\documentclass[aps,prd,onecolumn,showpacs,groupedaddress,nofootinbib]{revtex4-2}
\usepackage{graphicx}% Include figure files
\usepackage{dcolumn}% Align table columns on decimal point
\usepackage{bm}% bold math
\usepackage{color}
\usepackage{longtable}
\usepackage{supertabular}
\usepackage[T1]{fontenc}
\usepackage{epstopdf}
\usepackage{amsmath}
\usepackage{amssymb}
\usepackage{setspace}
\usepackage{multirow}
\usepackage{booktabs}
\usepackage[section]{placeins}
\usepackage{mathrsfs}
\usepackage{hyperref}
\usepackage{adjustbox}

\makeatletter

\newcommand{\Rmnum}[1]{\expandafter\@slowromancap\romannumeral #1@} 
\newcommand{\bq}{\begin{equation}}
\newcommand{\eq}{\end{equation}}
\newcommand{\bqn}{\begin{eqnarray}}
\newcommand{\eqn}{\end{eqnarray}}

\newcommand{\lb}{\label}

\makeatother

\begin{document}
\title{Sensitivity of black hole spectral instability to ultraviolet perturbations}

\author{Ramin G. Daghigh\textsuperscript{1}}\email[E-mail: ]{ramin.daghigh@metrostate.edu}
\author{Michael D. Green\textsuperscript{2}}\email[E-mail: ]{michael.green@metrostate.edu}
\author{Guan-Ru Li\textsuperscript{3}}
\author{Jodin C. Morey\textsuperscript{4}}\email[E-mail: ]{moreyj@lemoyne.edu}
\author{Wei-Liang Qian\textsuperscript{5,3,6}}\email[E-mail: ]{wlqian@usp.br}
\author{Stefan J. Randow\textsuperscript{7}}

\affiliation{$^{1}$ Natural Sciences Department, Metropolitan State University, Saint Paul, Minnesota, 55106, USA}
\affiliation{$^{2}$ Mathematics and Statistics Department, Metropolitan State University, Saint Paul, Minnesota, 55106, USA}
\affiliation{$^{3}$ Faculdade de Engenharia de Guaratinguet\'a, Universidade Estadual Paulista, 12516-410, Guaratinguet\'a, SP, Brazil}
\affiliation{$^4$ Le Moyne College, Syracuse, New York, 13214-1301, USA}
\affiliation{$^{5}$ Escola de Engenharia de Lorena, Universidade de S\~ao Paulo, 12602-810, Lorena, SP, Brazil}
\affiliation{$^{6}$ Center for Gravitation and Cosmology, College of Physical Science and Technology, Yangzhou University, Yangzhou 225009, China}
\affiliation{$^{7}$ Mathematics and Statistics Departmentent, University of Minnesota Duluth, Duluth, Minnesota, 55812, USA}

\begin{abstract}
Black hole quasinormal modes are known to exhibit spectral instability under ultraviolet perturbations of the effective potential.
In the present work, we investigate the sensitivity of the fundamental mode to different types of localized perturbations through a combination of analytic and numerical analyzes.
We show that the instability is governed primarily by the effective size of the perturbation rather than by its specific shape.
In particular, the instability may persist even in the limit where the width of the perturbation vanishes, provided that the integrated strength of the perturbation is not zero.
While a delta-function perturbation destabilizes the fundamental mode through an outward spiral, its interplay with a jump-discontinuity-type perturbation gives rise to competing inward and outward spiral motions.
We further show that the stability of the fundamental mode depends sensitively on how the magnitude of the perturbation decreases as it moves away from the compact object, leading to qualitatively distinct outward spirals, inward spirals, and rotational trajectories.
Finally, we investigate the motion of the fundamental mode in perturbed Regge-Wheeler potentials containing a jump discontinuity associated with a thin matter shell surrounding the black hole.
The resulting behavior qualitatively resembles the spiral structure observed in double-sided P\"oschl-Teller potentials, suggesting that the mechanisms identified in analytically tractable models persist in more realistic black hole effective potentials.
The present results indicate that the spectral instability of low-lying black hole modes is considerably richer than previously anticipated and may have important implications for black hole spectroscopy in realistic astrophysical environments.
\end{abstract}

%\pacs{ 03.75.Dg, 06.30.Gv, 37.25. + k, 91.10.Pp}

%\date{\today}
\date{July 24th, 2026}

\maketitle

\newpage
\section{Introduction}\label{sec1}

Black holes are one of the most captivating concepts in theoretical physics. 
They exemplify the properties of gravity at its extreme.  
The detections of gravitational waves by the LIGO and Virgo collaborations~\cite{agr-LIGO-01, agr-LIGO-02, agr-LIGO-03, agr-LIGO-04}, from binary black hole mergers, have opened a new window to the strong-field regime of gravity.  
Spectroscopy of the vibrational modes during the ringdown phase of a binary black hole merger can reveal important information about black holes and the relativistic regime of gravity.  

The natural vibrational modes, also called quasinormal modes (QNMs), of black holes suffer from spectral instability~\cite{spectral-instability-review-03, spectral-instability-review-05, spectral-instability-review-20} as a result of introducing a perturbation to the spacetime metric, which leads to a perturbed effective potential of the linearized wave equation equipped with the boundary conditions of an open system. 
Following the pioneering works of Nollert and Price~\cite{agr-qnm-instability-02, agr-qnm-instability-03} and of Aguirregabiria and Vishveshwara~\cite{agr-qnm-27, agr-qnm-30}, it has been shown~\cite{agr-qnm-instability-11, agr-qnm-Poschl-Teller-16, agr-qnm-echoes-20} that even tiny metric perturbations far from the compact object can dramatically affect the higher overtones.
These results challenge the intuitive expectation that small deformations of the effective potential would induce only minor changes in the QNM spectrum. In fact, the affected high-overtone modes tend to migrate toward the real frequency axis, in sharp contrast with the typical pattern where higher overtone modes have larger imaginary parts in most black hole backgrounds~\cite{agr-qnm-continued-fraction-12, agr-qnm-continued-fraction-23}.
Motivated by this behavior, Jaramillo {\it et al.}~\cite{agr-qnm-instability-07, agr-qnm-instability-13} systematically examined spectral instability in black hole perturbation theory using pseudospectrum techniques applied to deterministic and random perturbations~\cite{agr-qnm-instability-84} of the effective potential.
Their analyzes reveal that the boundary of the pseudospectrum moves toward the real frequency axis, leading to the picture of a universal instability of high-overtone modes triggered by {\it ultraviolet} (i.e., small-scale) perturbations~\cite{agr-qnm-instability-07, agr-qnm-instability-13}.
Along similar lines, many authors have explored related mechanisms~\cite{agr-bh-superradiance-27, agr-qnm-instability-44, agr-qnm-instability-75}, and it has been emphasized that such ultraviolet metric perturbations~\cite{agr-qnm-instability-07}, even when localized well away from the event horizon~\cite{agr-qnm-echoes-20} and seemingly physically insignificant~\cite{agr-qnm-instability-47}, can induce sizable deformations of the QNM spectrum.

The potential observational relevance of these findings lies in their impact on black hole spectroscopy~\cite{agr-BH-spectroscopy-19, agr-BH-spectroscopy-review-03, agr-BH-spectroscopy-review-04}.
The detections of gravitational waves from compact binary mergers~\cite{agr-LIGO-01, agr-LIGO-02, agr-LIGO-03, agr-LIGO-04} using ground-based facilities and the prospects of future spaceborne detectors, such as LISA~\cite{agr-LISA-01}, TianQin~\cite{agr-TianQin-01, agr-TianQin-Taiji-review-01}, and Taiji~\cite{agr-Taiji-01}, have fueled the expectation that precise measurements of ringdown waveforms may be achievable~\cite{agr-TianQin-05}.
In realistic astrophysical environments, however, the sources of gravitational waves are not isolated objects but are typically embedded in and interacting with surrounding matter, leading to deviations from ideal, highly symmetric metrics.
This has motivated extensive studies of ``dirty'' black holes~\cite{agr-bh-thermodynamics-12, agr-qnm-33, agr-qnm-34, agr-BH-spectroscopy-10}, whose spectral stability properties warrant careful scrutiny in view of the potential interpretation of observational data.
The deformation of the high overtone QNMs is intimately related to the phenomenon of gravitational-wave echoes, late-time ringing features first proposed by Cardoso {\it et al.}~\cite{agr-qnm-echoes-01, agr-qnm-echoes-review-01}.
As potential observables, echoes have been suggested as discriminators between otherwise similar gravitational systems through their distinct near-horizon properties. 
Through the analytic properties of the frequency-domain Green's function~\cite{agr-qnm-echoes-15, agr-qnm-echoes-16, agr-qnm-echoes-20, agr-strong-lensing-correlator-15, agr-qnm-echoes-45, agr-qnm-instability-65}, it has been argued that spectral instability is closely linked to echoes.
Broader studies on spectral instability, echoes, and causality~\cite{agr-qnm-instability-08, agr-qnm-instability-13, agr-qnm-instability-14, agr-qnm-instability-15, agr-qnm-instability-16, agr-qnm-instability-18, agr-qnm-instability-19, agr-qnm-instability-26, agr-qnm-echoes-22, agr-qnm-echoes-29, agr-qnm-echoes-30, agr-qnm-instability-23, agr-qnm-instability-29, agr-qnm-instability-32, agr-qnm-instability-33, agr-qnm-instability-43, agr-qnm-echoes-48, agr-qnm-echoes-35,agr-qnm-instability-70, agr-qnm-instability-71, agr-qnm-instability-87, agr-qnm-continued-fraction-40} have been carried out.
More recently, the notion of reflectionless scattering modes was elaborated in \cite{agr-qnm-instability-63, agr-qnm-instability-85, agr-qnm-echoes-50, agr-qnm-echoes-51, agr-qnm-echoes-52}, motivated by the role of greybody factors as potentially more robust observables in the presence of spectral instability~\cite{agr-qnm-instability-61, agr-qnm-instability-63, agr-qnm-Regge-14}.
In the high-frequency asymptotic regime, these modes are approximately uniformly distributed along the real frequency axis and have been shown to be primarily responsible for the echo phenomenon of the underlying compact object.
Another issue of concern is the stability of low-lying modes.
Cheung {\it et al.}~\cite{agr-qnm-instability-15} pointed out that even the fundamental mode of the Regge-Wheeler potential can appear unstable under generic perturbations, showing that a small deformation of the effective potential can drive the fundamental mode along an outward spiral in the complex-frequency plane.
In \cite{agr-qnm-instability-56, agr-qnm-instability-58}, this behavior was interpreted in an approximate setting where the background potential and its perturbation are modeled as disjoint barriers.
However, when this disjointness assumption is relaxed, a more subtle picture emerges. 
It was shown in \cite{agr-qnm-instability-55} that in a particular case the fundamental mode can, in fact, remain stable, a conclusion supported by fully numerical calculations without auxiliary approximations.
Nonetheless, these results reinforce the view that low-lying modes are significantly more robust than high overtones, even in the presence of ultraviolet perturbations, and highlight that the precise evaluation of QNMs as a result of these perturbations is essential for a reliable interpretation of ringdown signals.
In the time domain, it was recently shown~\cite{agr-qnm-instability-16} that black holes whose effective potential contains an insignificant bump exhibit ringdown signals with late-time echoes.
However, within the observable portion of the ringdown signal accessible to current detectors, QNMs affected by spectral instability have not been identified, despite attempts to extract them using various methods.  This is in agreement with previous results for black holes embedded in astrophysical environments~\cite{agr-BH-spectroscopy-60}.
The time-domain waveforms appear stable~\cite{agr-qnm-instability-83} when the Regge–Wheeler potential is approximated by a piecewise step-function profile, consistent with earlier studies~\cite{agr-qnm-instability-02, agr-qnm-instability-11}.
On the other hand, some preliminary results~\cite{agr-TDI-52} based on Bayesian analysis suggest that it may be feasible to assess the impact of these effects by varying different model parameters.
The extent to which such phenomena may pose additional challenges for current and future black hole spectroscopy programs remains an open question~\cite{agr-BH-spectroscopy-review-04}.

% The notion of spectral instability is by itself an intriguing idea that contradicts many intuitive assumptions of the properties of black hole QNMs.
% On the one hand, the effect of the perturbation, more specifically a perturbative barrier implemented on the effective potential, is shown to be even stronger as it moves farther away from the black hole~\cite{agr-qnm-instability-15}.
% On the other hand, at certain limits, it has been shown that the degree of instability vanishes concurrently as the magnitude of the perturbation decreases.
% In particular, several authors have scrutinized the instability in the fundamental mode in a disjointed effective potential~\cite{agr-qnm-instability-55, agr-qnm-echoes-22, agr-qnm-instability-15, agr-qnm-instability-65}.
% For the sake of mathematical simplicity, one can assume that the perturbation is implemented on the effective potential in the form of a rectangular barrier.
% It was argued in \cite{agr-qnm-instability-55, agr-qnm-instability-65} that as the width of the rectangular barrier vanishes, while maintaining its height, the instability of the fundamental mode disappears.
% The latter coincides with a vanishing reflection amplitude, along with the size of the perturbative barrier.
% At first glance, this result seems rather intuitive, as a barrier with a vanishing width will not interfere with wave propagation.

The present study is motivated by previous work and aims to further investigate the sensitivity of QNMs to ultraviolet metric perturbations.
Since the fundamental mode has a potentially more sizable impact on time-domain waveforms than the higher overtones, our discussion will be devoted primarily to the former.
One of our focuses is the specific shape of the perturbation. 
As expected, the degree of instability decreases with the magnitude of the perturbations~\cite{agr-qnm-instability-55, agr-qnm-instability-84}.
Regarding the width of the bump, one might assume, for the sake of mathematical simplicity, that the perturbation is implemented on the effective potential in the form of a rectangular bump.
It can be shown~\cite{agr-qnm-instability-55, agr-qnm-instability-65} that as the width of the bump vanishes, while maintaining its height, the instability of the fundamental mode disappears.
The latter coincides with a vanishing reflection amplitude, along with the size of the perturbative bump.
This result seems rather intuitive, as a bump with a vanishing width will not interfere with wave propagation.
It is worth pointing out that most studies on the instability of the low-lying modes argued that the behavior could be captured using a simplified scenario, where the perturbative bump is disjoint from the main black hole effective potential~\cite{agr-qnm-instability-50, agr-qnm-echoes-22, agr-qnm-instability-15, agr-qnm-instability-65}.
However, further analyses have indicated that a carefully placed perturbation on a continuous effective potential might significantly change the outcome~\cite{agr-qnm-instability-55, agr-qnm-instability-87}.

Another intriguing issue concerns the physical characterization of the strength and the location of the perturbation. 
In~\cite{agr-qnm-instability-59}, Boyanov argued that the strength of a localized perturbative bump in the effective potential is more appropriately characterized by its energy norm, which effectively measures the integrated strength of the perturbation, rather than simply by its peak amplitude or its width. In particular, in relation to the width of the bump, we show that in the extreme limit where the width approaches zero, the instability of the fundamental mode may nevertheless persist.
Physically, the energy norm reflects the total energy stored in the perturbative bump and therefore provides a more meaningful characterization of its overall strength. 
From a mathematical perspective, the influence of such perturbations on the QNM spectrum is naturally described by the $\epsilon$-pseudospectrum defined with respect to the energy scalar product~\cite{agr-qnm-instability-08}. 
For perturbations with fixed peak amplitude and other parameters, the corresponding energy norm increases as $r^{2}$, where $r$ denotes the radial coordinate. 
Consequently, displacing the perturbation farther from the compact object while keeping its peak amplitude fixed does not preserve its physical strength, since the integrated strength of the perturbation increases with distance. 
This scaling provides a natural explanation for the enhanced spectral instability observed in such cases. 
By contrast, when the perturbation is normalized so that the area under the bump scales as $1/r^{2}$, thereby maintaining an approximately constant energy norm, the apparent spectral instability is found to be substantially reduced for both the modified P\"oschl-Teller potential~\cite{agr-qnm-instability-84} and the Regge-Wheeler potential~\cite{agr-qnm-instability-87}. 
Throughout this work, we therefore consider both normalized and unnormalized perturbations in order to distinguish the intrinsic dependence of the QNM spectrum on the perturbation location from the effect of varying the perturbation's physical strength.

Although the energy norm provides a more physically meaningful characterization of a perturbative bump, it does not fully determine the spectral response to a more general perturbation. The underlying picture is more intricate and depends on the nature of the perturbation and the shape of the effective potential.
% Although the energy norm provides a more physically meaningful characterization of the perturbation strength, it does not fully determine the spectral response. The underlying picture is more intricate and also depends on other properties of the perturbative bump. 
% In particular, in relation to the width of the perturbative bump, we show that in the extreme limit where the width approaches zero, the instability of the fundamental mode may nevertheless persist.
% This observation suggests that, again, a more appropriate criterion for suppressing the instability is furnished by reducing  the area under the perturbative bump.
In addition to numerical evidence, we support this conclusion through analytic assessment.
% We then quantify the impact of the area under the bump on both the onset of instability and the spiral trajectory of the fundamental mode as the perturbation is displaced relative to the central black hole.
Several intriguing features associated with the stability of the fundamental mode are identified.
In particular, we present an example in which the mode initially spirals outward from one fundamental branch and subsequently spirals inward toward another.
The resulting spiral structure becomes even more intricate in the case of the Regge-Wheeler potential, which we investigate numerically for the first time.
While our analysis is not intended to be exhaustive, it aims to provide a systematic overview of the sensitivity of black hole spectra to such perturbations, as well as to discuss the associated observational implications.

% Intriguing enough, as will be elaborated in the present study, we argue that the situation is more complex.  For example, we show that in the extreme case of a delta-function barrier, where the width is zero,  the instability of the fundamental mode persists.  For the instability of the fundamental mode to disappear, it is more precise to say that the area of the perturbative barrier should vanish.  In addition to the numerical calculations, we support this idea with analytic results.    
% We also explore how decreasing the size of the barrier as it moves away from the central black hole impacts the
% spiral. We show that the stability of the fundamental mode depends on the rate of decrease of the barrier size.
% We then proceed to elaborate on a few more examples in which the phenomenon displays some interesting features, including outspiraling from one fundamental mode and subsequently inspiraling toward another.  The spiral behavior of the fundamental mode is even more complex for the Regge-Wheeler potential, which we explore numerically for the first time. 
% Even though the present study does not attempt to enumerate all possibilities, we strive to give a broad account of the sensitivity of black hole spectral instability and its observational implications.

In this study, we consider the setup in which the dynamics of black hole perturbations can be reduced to a master equation~\cite{agr-qnm-review-03,agr-qnm-review-06},
\begin{eqnarray}
\frac{\partial^2}{\partial t^2}\Psi(t, x)+\left(-\frac{\partial^2}{\partial x^2}+V_\mathrm{eff}\right)\Psi(t, x)=0 ,
\label{master_eq_ns}
\end{eqnarray}
with $x$ representing the tortoise coordinate, which is related to the radial coordinate according to $dx=dr/f(r)$, where $f(r)$ is the metric function defined below.
The effective potential $V_\mathrm{eff}$ is determined by the given spacetime metric, spin $s$, and angular momentum $\ell$ of the perturbation.
For instance, the Regge-Wheeler potential for a Schwarzschild black hole with mass $M$ reads
\bqn
V_\mathrm{RW}=f(r)\left[\frac{\ell(\ell+1)}{r^2}+(1-s^2)\frac{r_h}{r^3}\right],
\label{V_RW}
\eqn
where $r_h=2M$ is the event horizon radius and
\bqn
f(r)=1-\frac{r_h}{r} .
\label{f_master}
\eqn
Another example is the P\"oschl-Teller effective potential given by
\begin{eqnarray}
{V}_\mathrm{PT}=\frac{V}{\cosh ^2(\kappa x)} ,
\label{V_PT}
\end{eqnarray}
where the parameters $\kappa$ and $V$ measure the shape of the potential. 

In the study of black hole spectral instability, one typically considers an effective potential of the form
\bqn
V_\mathrm{eff} = V_\mathrm{BH} + V_\mathrm{pert} ,\label{ModVeff}
\eqn
where the unperturbed background potential $V_\mathrm{BH}$ can be given by Eq.~\eqref{V_RW} or~\eqref{V_PT}, and $V_\mathrm{pert}$ represents a perturbative bump located far from the compact object.

Throughout this work, we consider localized perturbations $V_\mathrm{pert}=V_\mathrm{pert}(r)$ that preserve the spherical symmetry of the underlying black hole spacetime. 
Such perturbations can be represented by radial modifications of the one-dimensional effective potential in the corresponding master equation. 
This constitutes an idealized setup intended to investigate the influence of localized ultraviolet metric perturbations on the quasinormal spectrum. 
In realistic astrophysical environments, perturbations are generally not spherically symmetric and may originate from localized objects, accretion flows, or other asymmetric matter distributions. 
The present model should therefore be regarded as describing the spherically symmetric sector, or equivalently the radial component, of a more general perturbation. 
Nevertheless, following the strategy of how the master equation~\eqref{master_eq_ns} is derived, linear perturbations on a spherically symmetric background admit a decomposition into independent angular multipoles.
Therefore, the mechanism identified here for a specific angular sector is expected to remain relevant whenever the localized perturbation can be treated perturbatively with the separation of variables being valid.

The quasinormal frequencies can be obtained by evaluating the zeros of the Wronskian
\begin{eqnarray}
W(\omega)\equiv W(g,h)=g(\omega,x)h'(\omega, x)-h(\omega,x)g'(\omega,x) ,
\label{pt_Wronskian}
\end{eqnarray}
where $'\equiv d/dx$, and $g$ and $h$ are the solutions of the corresponding homogeneous equation \eqref{master_eq_ns} in the frequency domain~\cite{agr-qnm-review-02},
\begin{eqnarray}
\left[-\omega^2-\frac{d^2}{dx^2}+V_\mathrm{eff}\right]\widetilde{\Psi}(\omega, x)=0 ,
\label{pt_homo_eq}
\end{eqnarray}
with appropriate boundary conditions, namely,
\begin{eqnarray}
\begin{array}{cc}
h(\omega, x)\sim e^{-i\omega x}    &~~~~~  x\to -\infty  \cr\\
g(\omega, x)\sim e^{i\omega x}     &~~~~~  x\to +\infty  
\end{array} .
\label{pt_boundary}
\end{eqnarray}
As discussed above, the present paper involves the choice of physically relevant effective potentials $V_\mathrm{eff}$ and their implication for the instability of the low-lying QNMs.

The remainder of the paper is organized as follows.
In Sec.~\ref{sec2}, we analyze an extreme limit in which the specific form of the perturbative bump becomes irrelevant as its width approaches zero.
To this end, we investigate several scenarios, with particular emphasis on the impact of a delta-function perturbation on the fundamental QNM for both continuous and discontinuous background effective potentials.
In particular, we elaborate on the competition between two types of ultraviolet perturbations, which leads to an interplay between inward and outward spirals.
In Sec.~\ref{sec3}, we examine the roles played by both the shape and the magnitude of the perturbative bump in determining the stability of the fundamental mode.
Specifically, we consider several scenarios in which the magnitude of the perturbative bump gradually decreases as it moves away from the compact object.
Another intriguing example involving inspiral and outspiral behavior is presented in Sec.~\ref{sec6}, arising from a P\"oschl-Teller potential with a jump discontinuity.
In Sec.~\ref{sec7}, we turn to more realistic scenarios involving the Regge-Wheeler effective potential subject to ultraviolet perturbations that mimic an infinitely thin matter shell wrapped around the central black hole.
Finally, concluding remarks are given in Sec.~\ref{secConclude}.

\section{The impact of an ultraviolet delta-function bump}\label{sec2}

In this section, we analyze an extreme limit in which the specific shape of the perturbative bump is not important.
This is relevant when the width of the bump vanishes faster than the remaining dimensions.
To this end, we explore the impact of a delta-function perturbative bump on the QNM spectrum of a black hole spacetime, where we effectively have a bump with zero width.  

%Let us introduce a deformed effective potential of the generic form  
Specifically, let us consider a perturbation to the effective potential
\bqn
V_\mathrm{pert} = \epsilon \delta(x-L) ,
\label{eq: deltaPert}
\eqn
% \bqn
% V_\mathrm{eff}=\widetilde{V}=
% \left\{\begin{array}{cc}
% V_l(x)     &  x < L  \cr\\
% \epsilon \delta(x-L)   &  x = L \cr\\ 
% V_r(x)     &  x > L 
% \end{array}\right. ,
% \label{}
% \eqn
%where the subscripts $l$ and $r$ refer to the left and right sides of the location of the delta-function at $L$ respectively.
where one places a delta-function at $x=L$.

As a comparison, consider the case in which the effective potential is discontinuous at $x=L$, but the discontinuity is less singular than a delta function (meaning it remains finite at $L$). 
In this case, the QNM spectrum is determined by the standard junction condition
\bqn
\frac{h'(L)}{h(L)}=\frac{g'(L)}{g(L)} ,
\label{eq: junction}
\eqn
which follows from integrating Eq.~\eqref{pt_homo_eq} across an infinitesimal interval containing the discontinuity, namely from $L-0$ to $L+0$.
If the effective potential remains finite at $x=L$, both the wave function and its first derivative are continuous across the discontinuity. 
The above condition is therefore equivalent to the vanishing of the Wronskian, evaluated at $x=L$,
\bqn
W(g,h)=g(L)h'(L)-h(L)g'(L) .
\lb{eqWronskianCond}
\eqn

The situation is different when the effective potential contains a delta-function contribution. 
Integrating Eq.~\eqref{pt_homo_eq} across the discontinuity still implies that the wave function itself is continuous,
\bqn
h(L)=g(L),
\eqn
whereas its first derivative exhibits a finite jump. More explicitly,
\bqn
h'(L)=g'(L)-\epsilon\, g(L),
\eqn
where the size of the jump is determined by the integrated singular part of the perturbation,
\bqn
\epsilon=\int_{L-0}^{L+0}V_{\mathrm{pert}}(x)\,dx.
\eqn

We then reorganize the above junction condition into
\bqn
\frac{h'(L)}{h(L)}=\frac{g'(L)-\epsilon g(L)}{g(L)},
\eqn
which leads to a modified Wronskian condition of the following form
\bqn
\bar{W}(g, h)\equiv W(g, h)+\epsilon h(L)g(L)={g}{h'}-{h}{g'}+\epsilon h g=0 .
\lb{modWronkianCond}
\eqn
The above is bilinear in $h$ and $g$.

\subsection{Outward spiral in a continuous effective potential with a delta-function perturbation}\label{sec31}

We consider a specific example in which a delta-function perturbation is superimposed onto a continuous P\"oschl-Teller potential, namely,
\bqn
V_\mathrm{BH}=V_\mathrm{PT} .
\eqn
Subsequently, the effective potential has the form
\bqn
V_\mathrm{eff}=
\left\{\begin{array}{cc}
{V}_\mathrm{PT}(x)     &  x < L  \cr\\
\epsilon \delta(x-L)   &  x = L \cr\\
{V}_\mathrm{PT}(x)     &  x > L
\end{array}\right. ,
\label{Eq:V_mpt}
\eqn
where $\epsilon$ is a parameter that controls the size of the delta-function.  
The QNM spectrum for the P\"oschl-Teller potential~\eqref{V_PT} can be found analytically, which is
\begin{eqnarray}
\omega_n=\sqrt{V-\frac{\kappa^2}{4}} - i \kappa \left(n+\frac{1}{2}\right),~~~n=0,1,2,\dots .
\label{V_PT-QNM}
\end{eqnarray}

The functions $h(x)$ and $g(x)$ in the modified Wronskian \eqref{modWronkianCond}, for the potential given in Eq. \eqref{Eq:V_mpt}, are
\bqn
h(x)=e^{-i \omega x} \left(e^{2 \kappa  x}+1\right)^{\beta } \, _2F_1\left(\beta ,\beta -\frac{i \omega}{\kappa };1-\frac{i \omega}{\kappa };-e^{2 \kappa  x}\right)
\label{eq: f-PT-delta}
\eqn
and
\bqn
g(x)=e^{i \omega x} \left(e^{-2 \kappa  x}+1\right)^{\beta } \, _2F_1\left(\beta ,\beta -\frac{i \omega}{\kappa };1-\frac{i \omega}{\kappa };-e^{-2 \kappa  x}\right),
\label{eq: g-PT-delta}
\eqn
where 
\bqn
\beta=\frac{1}{2}\left(1+\sqrt{1-4\frac{V}{\kappa^2}}\right)
\eqn
and $\, _2F_1\left(a,b;c;z\right)$ is the ordinary hypergeometric function. 
For large values of $z=e^{2 \kappa  x}$, we can expand the hypergeometric functions that appear in the modified Wronskian \eqref{modWronkianCond} using  
\bqn
\, _2F_1(a,b;c;-z)&=&z^{-a} \left(\frac{\Gamma (b-a) \Gamma (c)}{\Gamma (b) \Gamma (c-a)}-\frac{a \Gamma (-a+b-1) \Gamma (c)}{\Gamma (b) \Gamma (-a+c-1) z}+O\left(z^{-2}\right)\right) \nonumber \\
&&+z^{-b} \left(\frac{\Gamma (a-b) \Gamma (c)}{\Gamma (a) \Gamma (c-b)}-\frac{b \Gamma (a-b-1) \Gamma (c)}{\Gamma (a) \Gamma (-b+c-1) z}+O\left(z^{-2}\right)\right),
\label{eq: Series1}
\eqn
and
\bqn
\, _2F_1\left(a,b;c;-\frac{1}{z}\right)=1-\frac{a b}{c z}+O\left(z^{-2}\right).
\label{eq: Series2}
\eqn
%\bqn
%&&\, _2F_1\left(\beta ,\beta -\frac{i \omega}{\kappa };1-\frac{i w}{\kappa };-z\right) \sim \nonumber \\
%&&~~~~~~~~~ z^{-\beta } \left\{z^{\frac{i \omega}{\kappa }} \left[\frac{\Gamma \left(1-\frac{i \omega}{\kappa }\right) \Gamma \left(\frac{i \omega}{\kappa }\right)}{\Gamma (1-\beta ) \Gamma (\beta )}+\mathcal{O}\left(z^{-1}\right)\right]+\left[\frac{\Gamma \left(1-\frac{i \omega}{\kappa }\right) \Gamma \left(-\frac{i \omega}{\kappa }\right)}{\Gamma \left(-\frac{i \omega}{\kappa }-\beta +1\right) \Gamma \left(\beta -\frac{i \omega}{\kappa }\right)}+\mathcal{O}\left(z^{-1}\right)\right]\right\}.
%\label{eq: Series1}
%\eqn
%Note that the approximation $g(x)\sim e^{i\omega x}$ is equivalent to truncating the potential to the right of the perturbation.
After expanding the Wronskian \eqref{modWronkianCond} to the leading order, using the above expressions, we arrive at
%\bqn
%\frac{[2\omega (1+e^{2\kappa L})+i(\epsilon+e^{2\kappa L}\epsilon-2\beta \kappa)]
%\Gamma\left(1-i\frac{\omega}{\kappa}\right)\Gamma\left(-i\frac{\omega}{\kappa}\right)}{\Gamma\left(1-\beta-i\frac{\omega}{\kappa}\right) \Gamma\left(\beta-i\frac{\omega}{\kappa}\right)} \nonumber \\ 
%-\sin (\pi  \beta ) e^{2 i \omega L} \text{csch}\left(\frac{\pi  \omega}{\kappa }\right) \left(2 \beta  \kappa -\epsilon  e^{2 \kappa  L}-\epsilon \right)=0.
%\eqn
%For large $L$, this can further be simplified to
\bqn
\frac{\kappa  (2 \omega +i \epsilon )}{\omega  \Gamma \left(1-\beta -\frac{i \omega }{\kappa }\right) \Gamma \left(\beta -\frac{i \omega }{\kappa }\right)}+\frac{\epsilon  \sin (\pi  \beta ) e^{2 i \omega L } \Gamma \left(\frac{i \omega }{\kappa }\right)}{\pi  \Gamma \left(1-\frac{i \omega }{\kappa }\right)}\approx 0.
\eqn
In the vicinity of the low-lying modes, where $\omega \sim \omega_n$, we find that
\bqn
-i \frac{n!}{\kappa} \delta \omega_n &\approx & \frac{1}{\Gamma \left(\beta -\frac{i \omega}{\kappa }\right)} \nonumber \\
%&&= \frac{ \sin (\pi  \beta ) \text{csch}\left(\frac{\pi  \omega_n}{\kappa }\right)  \Gamma \left(1-\beta -\frac{i \omega_n}{\kappa }\right) }{ \Gamma \left(1-\frac{i \omega_n}{\kappa }\right)^2 }\nonumber \\
%&& ~~~ \times \frac{\omega  \epsilon  (\kappa +i \omega ) \left(e^{2\kappa L} (\kappa -i \omega )-2 \beta  (\beta  \kappa -i \omega )\right)}{\kappa  \left(e^{2\kappa L} \left(\kappa ^2+\omega ^2\right) (\epsilon -2 i \omega )-2 \beta  \left(\beta  \kappa ^2 \epsilon +\omega  \left(\omega  (\epsilon -2 i \omega )-2 i \kappa ^2\right)\right)\right)}e^{2 i \omega_n L}\nonumber \\
&\approx & \frac{ \sin (\pi  \beta ) \text{csch}\left(\frac{\pi  \omega_n}{\kappa }\right)  \Gamma \left(1-\beta -\frac{i \omega_n}{\kappa }\right) }{ \Gamma \left(1-\frac{i \omega_n}{\kappa }\right)^2 }\frac{\omega_n  \epsilon  }{\kappa  (\epsilon-2 i \omega_n ) }   e^{2 i \omega_n L} ,
\label{Eq.:AnalyticSpiral}
\eqn
where $\delta \omega_n= \omega-\omega_n \ll  \kappa$.
The above result shows that, for large values of $L$,
% where $e^{2 \kappa  L}  \gg 2 |\beta| \kappa / \epsilon$, 
the QNMs $\omega_n$ spiral outward as $L$ increases, since $\delta \omega_n \propto e^{2 i \omega_n L}= e^{2 i L \Re \omega_n}e^{-2 L \Im \omega_n}$ and $\Im \omega_n < 0$.  See Fig.~\ref{fig: Spiral}.

\subsection{Inward and outward spiral in a discontinuous effective potential with a delta-function perturbation}\label{sec32}

As discussed in the Introduction, it was shown in~\cite{agr-qnm-instability-55} that, for the modified P\"oschl-Teller effective potential in which the perturbation is implemented in terms of a small jump discontinuity, the fundamental mode remains stable.
This setup is physically relevant because it mimics~\cite{agr-qnm-continued-fraction-40} an infinitely thin ``mass shell'' surrounding the compact object.
In light of the results obtained in the previous subsection, the delta-function and jump discontinuity, although both intrinsically localized and small-scale in nature, appear to produce the opposite outcomes.

\begin{figure}[th!]
	\begin{center}
        \includegraphics[height=8cm]{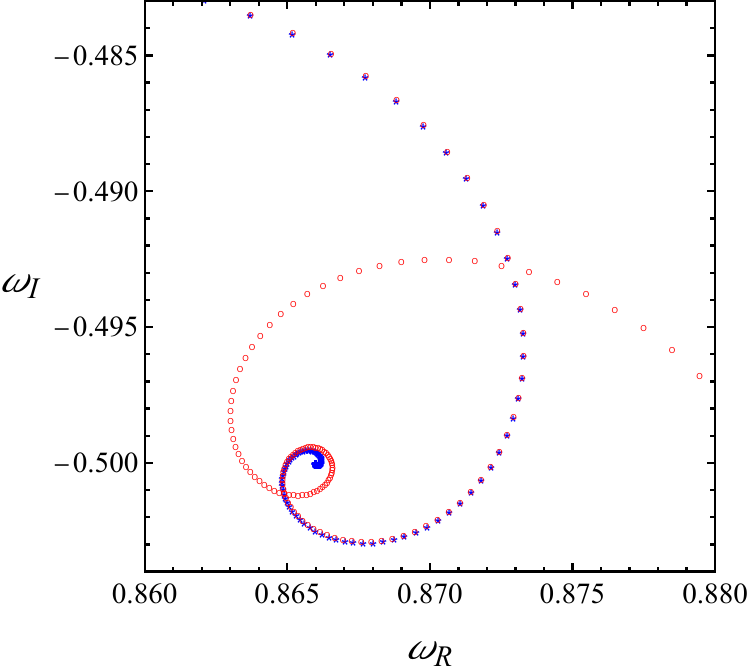}
	\end{center}
	\vspace{-0.7cm}
	\caption{\footnotesize 
    The outward spiral of the fundamental QNM in the frequency domain (blue stars) is plotted using the analytic result in Eq. \eqref{Eq.:AnalyticSpiral} for the potential \eqref{Eq:V_mpt} with the delta-function type perturbation, where we choose $V=1 \kappa^2$, and $\epsilon =10^{-5}\kappa$.   For comparison we show with red circles the inward, followed by an outward, spiral using the analytic result in Eq. \eqref{Eq.:AnalyticSpiral-withDisontinuity}, which was derived by introducing a jump discontinuity at $x=L$ to the effective potential \eqref{Eq:V_mpt}.  We use $\tilde{V}=0.5 \kappa^2$ in the region $x> L$.  The interval we use for $L$ is $[2.4\kappa^{-1},9.3\kappa^{-1}]$.}
	\label{fig: InOutSpiralDelta}
\end{figure}

Motivated by the above observation, we proceed to investigate a scenario in which both types of perturbations are simultaneously present.
This is achieved by introducing a jump discontinuity into the effective potential at the location of the delta function, $x=L$.
Specifically, the effective potential in Eq.~\eqref{Eq:V_mpt} is replaced, for the region $x>L$, by
\bqn
V_\mathrm{eff}(x>L)=\tilde{V}_\mathrm{PT}(x)
=\frac{\tilde{V}}{\cosh ^2(\kappa x)},
\eqn
where $\tilde{V}\neq V$.
In this case, the function $g(x)$ takes the new form
\bqn
\tilde{g}(x)=e^{i \omega x} \left(e^{-2 \kappa  x}+1\right)^{\tilde{\beta} } \, _2F_1\left(\tilde{\beta} ,\tilde{\beta} -\frac{i \omega}{\kappa };1-\frac{i \omega}{\kappa };-e^{-2 \kappa  x}\right),
\label{}
\eqn
where
\bqn
\tilde{\beta}=\frac{1}{2}\left(1+\sqrt{1-4\frac{\tilde{V}}{\kappa^2}}\right).
\eqn
Once again, we expand the Wronskian \eqref{modWronkianCond}, with $\tilde{g}$ replacing $g$, to the next-to-leading order for large $e^{2\kappa L}$ and arrive at
\bqn
-i \frac{n!}{\kappa} \delta \omega_n &\approx & \frac{1}{\Gamma \left(\beta -\frac{i \omega}{\kappa }\right)} \nonumber \\
&\approx & \frac{ \sin (\pi  \beta ) \text{csch}\left(\frac{\pi  \omega_n}{\kappa }\right)  \Gamma \left(1-\beta -\frac{i \omega_n}{\kappa }\right) }{ \Gamma \left(1-\frac{i \omega_n}{\kappa }\right)^2 }\frac{\omega_n  \epsilon  }{\kappa  (\epsilon-2 i \omega_n ) }   e^{2 i \omega_n L} \nonumber \\
&& +i \frac{\sin (\pi  \beta ) \text{csch}\left(\frac{\pi  \omega_n }{\kappa }\right) \Gamma \left(1-\beta -\frac{i \omega_n }{\kappa }\right) }{ \Gamma \left(1-\frac{i \omega_n }{\kappa }\right) \Gamma \left(2-\frac{i \omega_n }{\kappa }\right)} \nonumber \\
&& ~~\times \frac{\left\{\tilde{\beta } \left[2 \kappa ^2+i \omega_n  \epsilon-\kappa  \tilde{\beta } (2 \kappa +\epsilon ) \right] -\beta  \left[2 \kappa ^2-i \omega_n  \epsilon-\kappa  \beta  (2 \kappa -\epsilon ) \right]\right\}}{\kappa^2   (2 +i \epsilon/\omega_n)}e^{-2 \kappa L}e^{2 i \omega_n L}.
\label{Eq.:AnalyticSpiral-withDisontinuity}
\eqn
Note that we now have two terms.  
The first term is identical to Eq.~\eqref{Eq.:AnalyticSpiral}, which is proportional to $e^{2 i \omega_n L}$, and produces an outward spiral.  
However, the second term is proportional to $e^{-2\kappa L}e^{2 i \omega_n L}$ and produces an inward spiral motion for the fundamental mode as $L$ increases. 
The combination of these two terms produces an interesting behavior.
Since the first term is suppressed by the factor $\epsilon$, it is much smaller in size initially (for small $L$) compared to the second term, which is not suppressed by the smallness of the delta function.
As a result, there is an initial inward spiral followed by an outward spiral.  
In Fig.~\ref{fig: InOutSpiralDelta}, we demonstrate this feature by comparing the results in Eqs.~\eqref{Eq.:AnalyticSpiral} and \eqref{Eq.:AnalyticSpiral-withDisontinuity}.  
For smaller values of $L$, the second term in Eq.~\eqref{Eq.:AnalyticSpiral-withDisontinuity} is dominant and we see an inward spiral in  Fig.~\ref{fig: InOutSpiralDelta}.  
As $L$ increases, the second term becomes negligible compared to the first term and we see an outward spiral that matches well with the outward spiral produced using Eq.~\eqref{Eq.:AnalyticSpiral}.

%If we consider the case where the potential is truncated to the right of the delta-funtion ($x>L$), Eq.~\eqref{Eq.:AnalyticSpiral} also applies for intermediate values of $L$.    This is due to the fact that the derivation was done assuming that, for large enough values of $L$, one can ignore the potential after the delta-function.  For the intermediate values of $L$, initially $e^{2\kappa  L}  < 2|\beta| \kappa / \epsilon$.  In this case, the fundamental mode will begin spiraling inward ($\delta \omega_n \rightarrow 0$ as $e^{2\kappa  L}  \rightarrow 2|\beta| \kappa / \epsilon$) before spiraling outward; see Fig.\  \ref{fig: Spiral}.  This is not surprising because, for intermediate values of $L$, the truncation significantly modifies the shape of the potential and, consequently, the location of its fundamental mode.  

\subsection{Outward spiral in a disjoint effective potential with a delta-function perturbation}\label{sec33}

It is instructive to examine the physical origin of the spiral structure in the disjoint case.
To be specific, we investigate the effect of a delta-function-type perturbation that is spatially separated from the original black hole effective potential.
It is worth noting that this setup represents a simplified scenario that has already been extensively explored in the literature~\cite{agr-qnm-instability-56, agr-qnm-instability-58, agr-qnm-instability-50, agr-qnm-instability-55}. 

To this end, we use a rectangular barrier as a toy-model of a black hole potential.
A small delta-function perturbation is then introduced at a distant location to represent a perturbation to the main barrier.
We denote the height and width of the main barrier by $h$ and $\sigma$, respectively, and place it at the origin $x=0$.
The delta-function perturbation~\eqref{eq: deltaPert} is located at $x=L \gg \sigma$ and has strength $\epsilon\ll h$.
Specifically, the background effective potential is given by~\cite{agr-qnm-instability-15, agr-qnm-lq-matrix-06, agr-qnm-instability-18}
\begin{equation}
		V_\mathrm{BH} = \left\{ \begin{array}{ll}
			0 & x \le -\frac{\sigma}{2}~\\  \\
			h & -\frac{\sigma}{2} < x < \frac{\sigma}{2}~\\    \\
			0  & x\ge \frac{\sigma}{2} ~\\    
		\end{array}
		\right. ~ ,      
		\label{Vsquare}
\end{equation}
where $\sigma/2 < L$.
% \begin{equation}
% 		V(x) = \left\{ \begin{array}{ll}
% 			0 & x \le -\frac{\sigma}{2}~\\  \\
% 			h & -\frac{\sigma}{2} < x < \frac{\sigma}{2}~\\    \\
% 			0  & \frac{\sigma}{2} \le x < L~\\    \\
% 			\epsilon \delta(x-L)   &  x \ge L~
% 		\end{array}
% 		\right. ~.      
% 		\label{Vsquare}
% \end{equation}
According to \cite{agr-qnm-echoes-22}, the QNMs are governed by the roots of the Wronskian
\begin{equation}
		W_{1+2}=-2 i \omega [m_1^{++}(\omega) m_2^{++}(\omega)+ e^{2 i \omega L} m_1^{-+}(\omega)m_2^{+-}(\omega)] ,\label{eq: QNMs}
	\end{equation}
where  $m_1^{++}$, $m_2^{++}$, $m_1^{-+}$, and $m_2^{+-}$ are the matching coefficients defined in \cite{agr-qnm-echoes-22}.
For this potential, 
	\begin{eqnarray}
		m_1^{++}(\omega)&=& \frac{e^{i\sigma \omega}}{4 \omega \hat{\omega}}\left[e^{i\sigma \hat{\omega}}(\omega-\hat{\omega})^2-e^{-i\sigma \hat{\omega}}(\omega+\hat{\omega})^2\right] ,\nonumber  \\
		m_2^{++}(\omega)&=& 1-\frac{\epsilon}{2 i \omega} ,\nonumber  \\
		m_1^{-+}(\omega)&=& \frac{e^{-i\sigma \omega}}{2 \omega \hat{\omega}} i h \sin{(\sigma \hat{\omega})} ,\nonumber \\
		m_2^{+-}(\omega)&=& -\frac{\epsilon}{2 i \omega}  ,
		\label{eq: TransitMatrix2S}
	\end{eqnarray}
	where $\hat{\omega}=\sqrt{\omega^2-h}$. 
	For a perturbative delta-function, $m_2^{+-}(\omega)\to 0$  as $ \epsilon\to 0$, and one recovers the QNM spectrum of the unperturbed potential from the equation $m_1^{++}(\omega)=0$.

As shown in \cite{agr-qnm-instability-65}, one can then find the deviation from the low-lying modes $\omega_n$ ($\delta \omega_n =\omega-\omega_n$) using 
	\begin{equation}
		\delta \omega_n \approx - \frac{m_1^{-+}(\omega_n)m_2^{+-}(\omega_n)}{m_1^{++}{'}(\omega_n) m_2^{++}(\omega_n)} e^{2 i \omega_n L}= \frac{\epsilon  \hat{\omega}_n^2 e^{-2 i \sigma  \omega_n }}{2\sigma \omega_n  \hat{\omega}_n \left[\hat{\omega}_n+i \omega_n  \cot \left(\sigma \hat{\omega}_n\right)\right]-2i h }e^{2 i \omega_n L}~.
		\label{spiralEq}
	\end{equation}
This is an outward spiral in the complex plane as $L$ increases.  
As expected, this is qualitatively similar to the result we found in Eq.~\eqref{Eq.:AnalyticSpiral}, where we also have an outward spiral with increasing $L$.

\section{The impact of the shape and location of the perturbation}\label{sec3}

In this section, we explore the role played by the shape of the perturbative bump.
The analysis focuses on two aspects.
First, we examine to what extent the specific shape of the perturbation affects the stability of the QNMs.
Second, we investigate how the strength of the perturbation, viewed as a function of its location, influences the QNMs.
Unless otherwise specified, the perturbation strength is characterized by its peak amplitude. 
Sec.~\ref{sec5} revisits this choice by introducing an energy-norm normalization, allowing the effects of the perturbation location and its integrated strength to be examined separately.

\subsection{ Thin rectangular perturbation of a continuous effective potential}

In this subsection, we start with a scenario in which a rectangular perturbative potential barrier is superimposed on a continuous background metric.
This is implemented through a simplified model consisting of a P\"oschl-Teller potential supplemented by a localized rectangular bump,
\begin{equation}
		V(x) = \left\{ \begin{array}{lll}
			V_\mathrm{PT}(x)  &~~~~~& x \le L-\frac{\sigma}{2}~\\    \\
			V_\mathrm{PT}\left(L-\frac{\sigma}{2}\right)+\varepsilon &~~~~~& L-\frac{\sigma}{2} < x < L+\frac{\sigma}{2}~\\    \\
			V_\mathrm{PT}(x) &~~~~~& x \ge L+\frac{\sigma}{2}~
		\end{array}
		\right. ,    
		\label{eq: VPTsquare}
	\end{equation}
where $\varepsilon$ characterizes the magnitude of the perturbation and is taken to be constant at this moment.
The wavefunction for the regions $x \le L-\frac{\sigma}{2}$ and  $x \ge L+\frac{\sigma}{2}$  are given in Eq.\ \eqref{eq: f-PT-delta} and Eq.\ \eqref{eq: g-PT-delta} respectively.   The wave function in the rigion $L-\frac{\sigma}{2} < x < L+\frac{\sigma}{2}$ is a linear combination of $e^{\pm i \tilde{\omega}x}$, where $\tilde{\omega}=\sqrt{\omega^2-\left[V_\mathrm{PT}\left(L-\frac{\sigma}{2}\right)+\varepsilon\right]}$. We can derive the Wronskian by applying the junction conditions [similar to Eq.~\eqref{eq: junction}] at $x=L\pm \sigma/2$.  After equating the obtained Wronskian to zero, we find
%For large values of $e^{2\kappa L}$ we can assume $g(x)\sim e^{i\omega x}$, which is equivalent to truncating the potential to the right of the barrier. Thus for large values of  $L$, the Wronskian for the above potential takes the form
\bqn
&&\tilde{\omega } \, _2F_1\left(\beta ,\beta +\frac{i \omega }{\kappa };\frac{i \omega }{\kappa }+1;-e^{-2 \kappa L -\kappa  \sigma }\right) \nonumber \\
&&\times \left[2 i \beta  \kappa  \left(1+e^{2 i \sigma  \tilde{\omega }}\right) \, _2F_1\left(\beta +1,\beta -\frac{i \omega }{\kappa };1-\frac{i \omega }{\kappa };-e^{2 \kappa L -\kappa  \sigma }\right) \right. \nonumber \\
&& \left. ~~~~~+\frac{\left(\left(1+e^{2 i \sigma  \tilde{\omega }}\right) \left(\omega  e^{2 \kappa  L}+e^{\kappa  \sigma } (\omega -2 i \beta  \kappa )\right)-\tilde{\omega } \left(-1+e^{2 i \sigma  \tilde{\omega }}\right) \left(e^{\kappa  \sigma }+e^{2 \kappa  L}\right)\right) \, _2F_1\left(\beta ,\beta -\frac{i \omega }{\kappa };1-\frac{i \omega }{\kappa };-e^{2 \kappa L -\kappa  \sigma }\right)}{e^{\kappa  \sigma }+e^{2 \kappa  L}}\right] \nonumber \\
&&+\left[\left(\frac{2 \beta  \kappa }{e^{\kappa  (2 L+\sigma )}+1}-i \omega \right) \, _2F_1\left(\beta ,\beta +\frac{i \omega }{\kappa };\frac{i \omega }{\kappa }+1;-e^{-2 \kappa L -\kappa  \sigma }\right) \right. \nonumber \\
&&  \left. ~~~~~-\frac{2 \beta  \kappa  (\beta  \kappa +i \omega ) e^{-2 \kappa L -\kappa  \sigma } \, _2F_1\left(\beta +1,\beta +\frac{i \omega }{\kappa }+1;\frac{i \omega }{\kappa }+2;-e^{-2 \kappa L -\kappa  \sigma }\right)}{\kappa +i \omega }\right]  \nonumber \\
&& \times \left[2 \beta  \kappa  \left(-1+e^{2 i \sigma  \tilde{\omega }}\right) \, _2F_1\left(\beta +1,\beta -\frac{i \omega }{\kappa };1-\frac{i \omega }{\kappa };-e^{2 \kappa L -\kappa  \sigma }\right) \right. \nonumber \\
&& \left. ~~~~~+\frac{i \left(\tilde{\omega } \left(1+e^{2 i \sigma  \tilde{\omega }}\right) \left(e^{\kappa  \sigma }+e^{2 \kappa  L}\right)-\left(-1+e^{2 i \sigma  \tilde{\omega }}\right) \left(\omega  e^{2 \kappa  L}+e^{\kappa  \sigma } (\omega -2 i \beta  \kappa )\right)\right) \, _2F_1\left(\beta ,\beta -\frac{i \omega }{\kappa };1-\frac{i \omega }{\kappa };-e^{2 \kappa L -\kappa  \sigma }\right)}{e^{\kappa  \sigma }+e^{2 \kappa  L}}\right]\nonumber \\
&& =0,
\label{eq: W-PT-square}
\eqn
% \bqn
% && 2 i \beta  \kappa  \left(\left(\tilde{\omega}-\omega\right) e^{2 i \sigma  \tilde{\omega}}+\tilde{\omega}+\omega\right) \, _2F_1\left(\beta +1,\beta -\frac{i \omega}{\kappa };1-\frac{i \omega}{\kappa };-e^{2  \kappa L-\kappa  \sigma }\right)  \nonumber \\
% &&+\left(2 \tilde{\omega} \left(1+e^{2 i \sigma  \tilde{\omega}}\right) \left(\omega e^{2 \kappa  L}+e^{\kappa  \sigma } (\omega-i \beta  \kappa )\right)+\omega \left(1-e^{2 i \sigma  \tilde{\omega}}\right) \left(\omega e^{2 \kappa  L}+e^{\kappa  \sigma } (\omega-2 i \beta  \kappa )\right)+\tilde{\omega}^2 \left(1-e^{2 i \sigma  \tilde{\omega}}\right) \left(e^{\kappa  \sigma }+e^{2 \kappa  L}\right)\right) \nonumber \\
% &&\frac{\, _2F_1\left(\beta ,\beta -\frac{i \omega}{\kappa };1-\frac{i \omega}{\kappa };-e^{2 \kappa L-\kappa  \sigma }\right)}{e^{\kappa  \sigma }+e^{2 \kappa  L}}=0 ,
% \label{eq: W-PT-square}
% \eqn
which determines the QNM spectrum.  For large values of $z=e^{2 \kappa  L}$, we can use the Taylor expansions in Eqs.~\eqref{eq: Series1} and \eqref{eq: Series2} to obtain
% \bqn
% &&\, _2F_1\left(\beta +1,\beta -\frac{i \omega}{\kappa };1-\frac{i \omega}{\kappa };-z\right)=  \nonumber \\
% &&~~~~~~~~~~~~~~ z^{-\beta -1} \left\{z^{\frac{i w}{\kappa }} \left[\frac{\Gamma \left(1-\frac{i w}{\kappa }\right) \Gamma \left(\frac{i w}{\kappa }+1\right) z}{\Gamma (1-\beta ) \Gamma (\beta +1)}+\mathcal{O}\left(z^0\right)\right]+\left[\frac{\Gamma \left(-\frac{i w}{\kappa }-1\right) \Gamma \left(1-\frac{i w}{\kappa }\right)}{\Gamma \left(-\frac{i w}{\kappa }-\beta \right) \Gamma \left(\beta -\frac{i w}{\kappa }\right)}+\mathcal{O}\left(z^{-1}\right)\right]\right\}.
% \label{eq: series2}
% \eqn
% The series expansion of the other hypergeometric function in Eq.\ \eqref{eq: W-PT-square} is given in Eq.\ \eqref{eq: Series1}.
% After combining Eqs.\ \eqref{eq: Series1}, \eqref{eq: W-PT-square}, and \eqref{eq: series2}, in the limit where $\kappa \sigma \ll 1 \ll \kappa L$, to the lowest order, we find
\bqn
-i \frac{n!}{\kappa} \delta \omega_n \approx \frac{1}{\Gamma \left(\beta -\frac{i \omega_n}{\kappa }\right)} \approx
-\frac{ \sin (\pi  \beta ) \text{csch}\left(\frac{\pi  \omega_n}{\kappa }\right)  \Gamma \left(1-\beta -\frac{i \omega_n}{\kappa }\right) }{ \Gamma \left(1-\frac{i \omega_n}{\kappa }\right)^2}
\frac{\bar{\varepsilon}\omega  e^{-i \omega \sigma } }{\kappa  \left(2 i \omega  \tilde{\omega } \cot \left(\sigma  \tilde{\omega }\right)+\tilde{\omega }^2+\omega ^2\right)}e^{2 i \omega_n L},
\label{Eq.:AnalyticSpiralRectangle}
\eqn
where $\bar{\varepsilon} = V_\mathrm{PT}\left(L-\frac{\sigma}{2}\right)+\varepsilon$.  Note that the rectangular bump with a fixed area resembles a delta function in the limit $\sigma \rightarrow 0$.  In this limit, the r.h.s.~of Eq.\ \eqref{Eq.:AnalyticSpiralRectangle} approaches
\bqn
-\frac{ \sin (\pi  \beta ) \text{csch}\left(\frac{\pi  \omega_n}{\kappa }\right)  \Gamma \left(1-\beta -\frac{i \omega_n}{\kappa }\right) }{ \Gamma \left(1-\frac{i \omega_n}{\kappa }\right)^2}
\frac{\bar{\varepsilon}\sigma }{2 i \kappa  }e^{2 i \omega_n L}.
\label{Eq.:AnalyticSpiralRectangle-rhs}
\eqn
As expected, this result is almost identical to what we found for the delta-function case in Eq.\ \eqref{Eq.:AnalyticSpiral}.  More precisely, for $\epsilon \ll \omega_n$, the r.h.s.~of Eq.\ \eqref{Eq.:AnalyticSpiral} reduces to Eq.\ \eqref{Eq.:AnalyticSpiralRectangle-rhs}.

\begin{figure}[th!]
	\begin{center}
        \includegraphics[height=8cm]{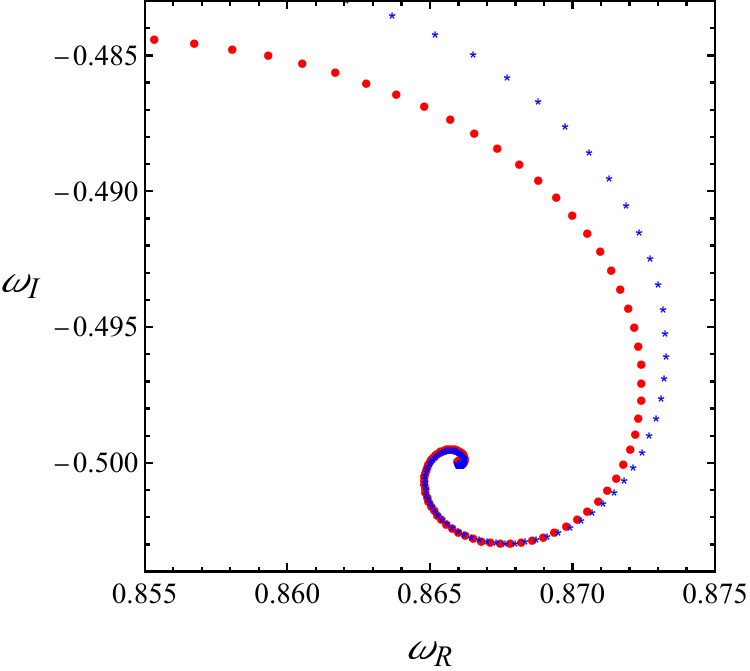}
	\end{center}
	\vspace{-0.7cm}
	\caption{\footnotesize 
    The outward spiral of the fundamental QNM in the frequency domain (blue stars) is plotted using the analytic result in Eq. \eqref{Eq.:AnalyticSpiral} for the potential \eqref{Eq:V_mpt} with the delta-function type perturbation, where we choose $V=1 \kappa^2$, and $\epsilon =10^{-5}\kappa$.   The same data points are also shown in Fig.~\ref{fig: InOutSpiralDelta} with blue stars.  In red dots, we show the numerical calculation done with no approximation for the potential in Eq.\ \eqref{eq: VPTsquare}, where the rectangular bump has a width of $ \sigma =10^{-4}\kappa^{-1}$ and a height of $\varepsilon=10^{-1} \kappa^2$. Note that we choose these values so that the area under the delta-function and the rectangular bump are approximately equal.  In this case the fundamental mode initially is very close to the fundamental mode of a pure P\"oschl-Teller potential, which is located at $\omega_0\approx(0.866-0.5 i) \kappa$ for this case.  The mode spirals out as the rectangular bump moves away from the central black hole.  The outward spirals of both cases match well near $\omega_0$.   The analytic approximation in Eq. \eqref{Eq.:AnalyticSpiral} becomes less reliable as the QNM moves away from $\omega_0$.  The analytic result in Eq.~\eqref{Eq.:AnalyticSpiralRectangle} generates almost exactly the same spiral produced by the Eq.~\eqref{Eq.:AnalyticSpiral} shown with the blue stars.}
    %The spiral of the fundamental QNM in the frequency domain (green stars) is plotted for the analytic result in Eq. \eqref{Eq.:AnalyticSpiral} for the potential \eqref{Eq:V_mpt} with the delta-function type perturbation, where we choose $V=1 \kappa^2$, and $\epsilon =10^{-5}\kappa$.  In blue circles, we show the same calculations done numerically for the Wronskian given in Eq.\ \eqref{eq: W-PT-square}, where the rectangular bump has a width of $ \sigma =10^{-4}\kappa$ and a height of $\varepsilon=10^{-1} \kappa^2$.  The results are  qualitatively similar.  The data points move from right to left as the perturbative bump moves away from the central black hole.   These results are valid for large values of $L$, in which one can assume the potential is truncated after the bump.   The inward spiral is a direct consequence of truncating the potential early, where $L$ is not large.  In red dots, we show the numerical calculation done with no approximation for the potential in Eq.\ \eqref{eq: VPTsquare}, where the rectangular bump has a width of $ \sigma =10^{-4}\kappa$ and a height of $\bar{\varepsilon}=10^{-1} \kappa^2$. In this case the fundamental mode initially is very close to the fundamental mode of a pure P\"oschl-Teller potential, which is located at $\omega_0\approx(0.866-0.5 i) \kappa$ for this case.  The mode spirals out as the rectangular bump moves away from the central black hole.  The outward spiral of all three cases match well near $\omega_0$.   The analytic result Eq. \eqref{Eq.:AnalyticSpiral} becomes less reliable as the QNM moves away from $\omega_0$.}
	\label{fig: Spiral}
\end{figure}

In Fig.\ \ref{fig: Spiral}, we show numerically, by solving Eq.\ \eqref{eq: W-PT-square}, that the fundamental mode for the effective potential in Eq.\ \eqref{eq: VPTsquare} spirals out from the location of the fundamental mode of the unperturbed effective potential $\omega_0$ as the perturbative bump moves away from the central black hole.  
In Fig.\ \ref{fig: Spiral}, We also show the spiral for the approximate analytic solution in Eq.~\eqref{Eq.:AnalyticSpiral}. When $\sigma\ll 1 \kappa$, the analytic results \eqref{Eq.:AnalyticSpiral} and \eqref{Eq.:AnalyticSpiralRectangle} have visibly indistinguishable spirals for the parameters used in Fig.\ \ref{fig: Spiral}.  The analytical and numerical results are in good agreement for small deviations $\delta\omega_n$.  As expected, analytical results become less reliable as the QNM moves away from the fundamental mode $\omega_0$, since analytical calculations are only valid for small $\delta\omega_n$.   
% The latter cases are qualitatively similar and involve an inward spiral prior to an outward spiral.  
% As mentioned earlier, these approximations are made assuming that the potential can be truncated for large values of $L$, where the height of the effective potential is small enough to be neglected.  In fact, these approximations match well with the numerical result for large values of $L$ near $\omega_0$.  For intermediate values of $L$, the truncation has a large impact on the effective potential and consequently on the location of the fundamental mode, which will no longer be close to $\omega_0$. Therefore, for an effective potential that is truncated right after the perturbative barrier, the fundamental mode presents an initial inspiral toward $\omega_0$ before it spirals out.  Note that in a truncated P\"oschl-Teller potential, with no perturbative barrier, the fundamental mode is stable and one only observes an inspiral toward $\omega_0$ \cite{agr-qnm-instability-55}.  Therefore, the presence of the perturbative barrier is necessary for the instability of the fundamental mode.   

\subsection{Non-rectangular perturbations of a continuous effective potential}\label{sec4}

In addition to the rectangular bump introduced in Eq.~\eqref{eq: VPTsquare}, we also consider perturbative bumps with a triangular and raised P\"oschl-Teller profiles.
The triangular-bump case consists of a P\"oschl-Teller potential supplemented by a piecewise linear perturbative bump of the form
\begin{equation}
		V(x) = \left\{ \begin{array}{lll}
			V_\mathrm{PT}(x)  &~~~~~& x \le L ~\\  \\
			V_\mathrm{PT}(L)+\dfrac{\varepsilon}{\delta} (x-L)   &~~~~~& L < x < L+\delta~\\    \\
			V_\mathrm{PT}(x) &~~~~~& x \ge L+\delta~
		\end{array}
		\right. ,    
		\label{Eq: TriangularBump}
	\end{equation}
where $\varepsilon$ is the height and $\delta$ is the width of the triangular bump.  The solution to the wave equation in the bump region is:
\bqn
C\cdot \text{Ai}\left[ \frac{-\omega^2 +V_\mathrm{PT}(L)+\dfrac{\varepsilon}{\delta} (x-L) }{\left(\varepsilon/\delta\right)^{2/3}} \right]
+D\cdot \text{Bi}\left[ \frac{-\omega^2 +V_\mathrm{PT}(L)+\dfrac{\varepsilon}{\delta} (x-L) }{\left(\varepsilon/\delta\right)^{2/3}} \right],
\label{}
\eqn
where $C$ and $D$ are constants, and $\text{Ai}(z)$ and $\text{Bi}(z)$ are the two linearly independent Airy functions.

Another case we consider is the combination of a P\"oschl-Teller potential with a raised P\"oschl-Teller perturbative bump of the form
\begin{equation}
		V(x) = \left\{ \begin{array}{lll}
			V_\mathrm{PT}(x)  &~~~~~& x \le L ~\\  \\
			V_\mathrm{PT}(x)+\varepsilon   &~~~~~& L < x < L+\sigma~\\    \\
			V_\mathrm{PT}(x) &~~~~~& x \ge L+\sigma~
		\end{array}
		\right. ,    
		\label{}
	\end{equation}
where, as usual, $\varepsilon$ is the height and $\sigma$ is the width of the bump. 

\begin{figure}[th!]
	\begin{center}
		\includegraphics[height=9cm]{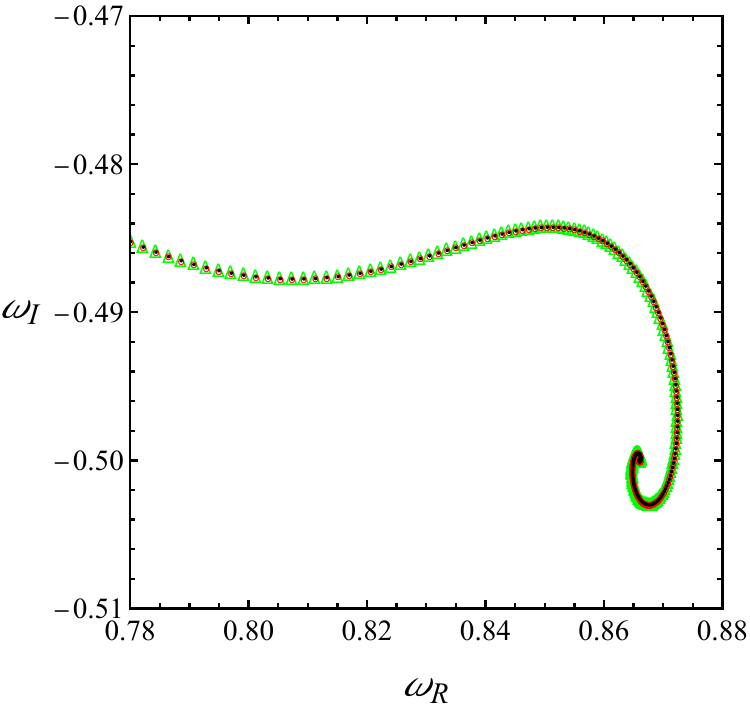}
	\end{center}
	\vspace{-0.7cm}
	\caption{\footnotesize The outward spiral of the fundamental QNM in the frequency domain is plotted for the P\"oschl-Teller potential with a rectangular bump (red circles), triangular bump (green triangles),  and raised P\"oschl-Teller bump (black dots).  There is almost no difference. We choose  $V=1 \kappa^2$,  $\varepsilon=10^{-1} \kappa^2$, and $\sigma =10^{-4}\kappa^{-1}$.  For the  triangular case, to keep the area under the bump the same as the rectangular bump, we choose  $\varepsilon=2\times 10^{-1} \kappa^2$ and $\delta= 10^{-4}\kappa^{-1}$.}
	\label{Fig: spiralAltBumps}
\end{figure}

\begin{figure}[th!]
	\begin{center}
		\includegraphics[height=9cm]{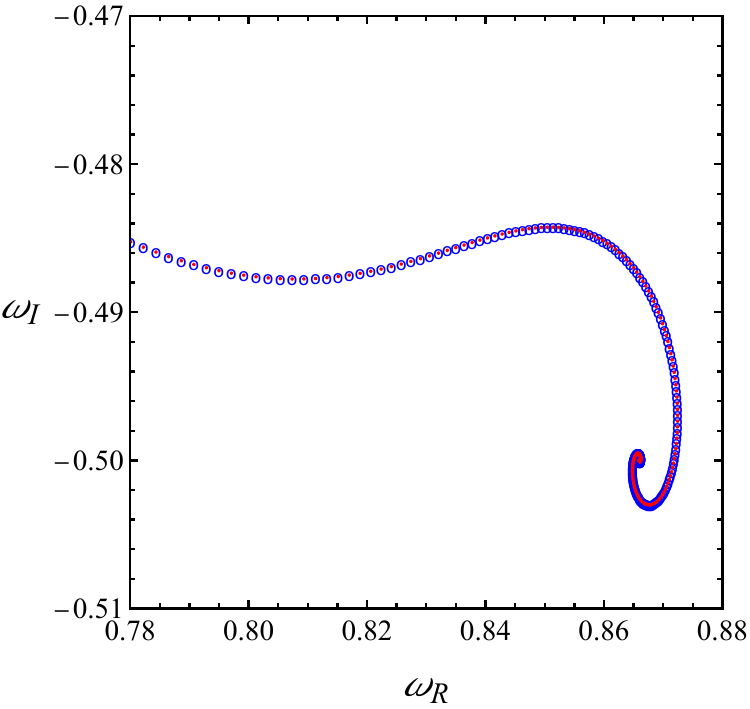}
	\end{center}
	\vspace{-0.7cm}
	\caption{\footnotesize The outward spiral of the fundamental QNM in the frequency domain is plotted for the P\"oschl-Teller potential with two different triangular bumps, Eq.~\eqref{Eq: TriangularBump}, which have the same area but different heights and widths.  We choose  $V=1 \kappa^2$.  The blue circles correspond to a triangle with  $\varepsilon=2\times 10^{-1} \kappa^2$, and $\delta =10^{-4}\kappa^{-1}$. The red dots correspond to a triangle with the same area, but with $\varepsilon=6\times 10^{-1} \kappa^2$, and $\delta =\frac{1}{3} \times 10^{-4}\kappa^{-1}$. The resulting spiral is the same for both cases indicating that the area under the perturbation, rather than the height, is the determining factor for the shape of the spiral.}
	\label{Fig: spiralAltTriangle}
\end{figure}

In Fig.\ \ref{Fig: spiralAltBumps}, we show the spiral of the fundamental QNM in the frequency domain for the P\"oschl-Teller potential with a rectangular bump, triangular bump, and raised P\"oschl-Teller bump.  There is almost no difference between the three cases.  In Fig.~\ref{Fig: spiralAltTriangle}, we show the behavior of the fundamental QNM for the P\"oschl-Teller potential with two different triangular bumps, which have the same area but different heights and widths.  Again, changing the height and width of the triangle does not affect the behavior of the fundamental mode as long as the area stays the same.
Therefore, the specific shape of the perturbative bump seems to be irrelevant to the instability of the fundamental mode. 

\subsection{The magnitude of the perturbative bump that diminishes as it moves away}\label{sec5}

We now turn to the second aspect by considering a few scenarios in which the magnitude of the perturbative bump gradually decreases as it moves away from the compact object.
 In Sec.~\ref{sec32}, the  discontinuous P\"oschl-Teller potential with a delta-function perturbation exhibited an inward and outward spiral.  The inward spiral was due to a term proportional to $e^{-2\kappa L}$.
In the following section, Sec.~\ref{sec6}, we will examine the case of a discontinuous P\"oschl-Teller potential, without a delta-function, and show that the size of the perturbation decreases proportionally to $e^{-2\kappa L}$ as the perturbation moves away from the central black hole.  We will see that this also leads to an inward spiral.  This suggests that the fundamental mode can be stabilized if the size of the bump diminishes quickly enough as it moves away from the central black hole. 

\begin{figure}[th!]
	\begin{center}
		\includegraphics[height=8cm]{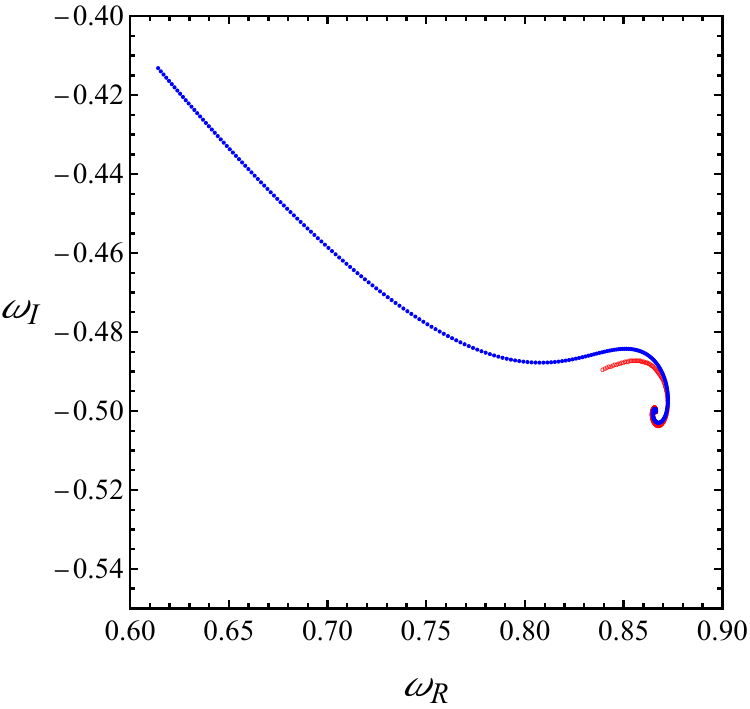}
	\end{center}
	\vspace{-0.7cm}
	\caption{\footnotesize The outward spiral of the fundamental QNM in the frequency domain is plotted, in blue dots, for the P\"oschl-Teller potential with a rectangular bump, given in Eq.~\eqref{eq: VPTsquare}, that moves away from the central black hole. We choose  $V=1 \kappa^2$,  $\varepsilon=10^{-1} \kappa^2$, and $\sigma =10^{-4}\kappa^{-1}$.  In red dots, we plot the same spiral with the difference that the height of the rectangular bump diminishes proportional to $1/L^2$.  More precisely, we have  $\varepsilon=\left(\frac{L_0}{L}\right)^2\varepsilon_0$, where we take $\varepsilon_0=10^{-1} \kappa^2$ and $L_0=2\kappa^{-1}$.}
	\label{Fig: spiral-1:x^2}
\end{figure}

\begin{figure}[th!]
	\begin{center}
		\includegraphics[height=9cm]{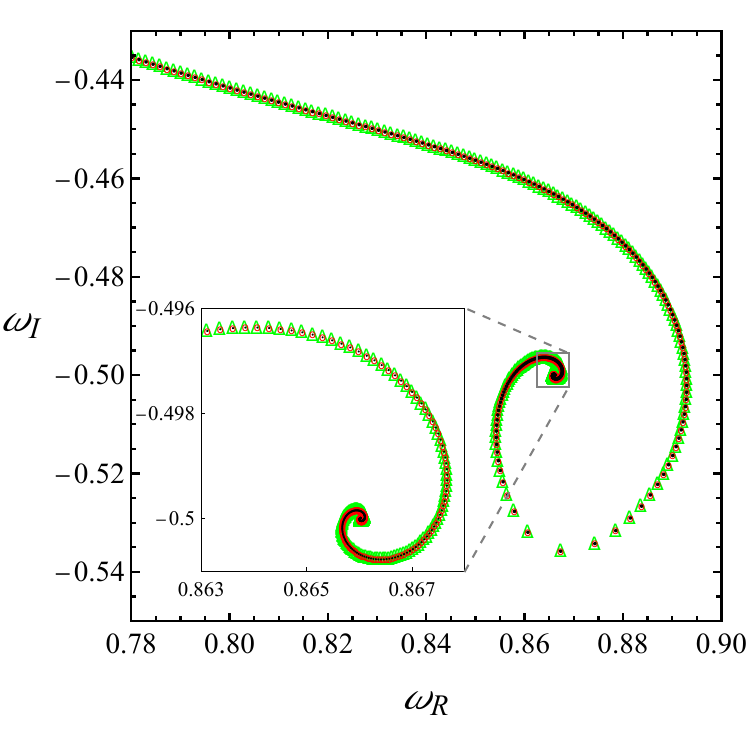}
	\end{center}
	\vspace{-0.7cm}
	\caption{\footnotesize We plot the same outward spiral of the fundamental mode as in Fig.~\ref{Fig: spiralAltBumps}, for the P\"oschl-Teller potential with a rectangular bump (red circles), triangular bump (green triangles),  and raised P\"oschl-Teller bump (black dots), with the difference that the height of the bump diminishes proportional to $1/L$.  More precisely, we have  $\varepsilon=\left(\frac{L_0}{L}\right)\varepsilon_0$, where we take $\varepsilon_0=10^{-1} \kappa^2$ and $L_0=2\kappa^{-1}$.  
    The number of data points in the inset panel is approximately the same as in Fig.~\ref{Fig: spiralAltBumps}.  
    This shows that the fundamental mode is less perturbed in this scenario.  Just as before, the perturbation of the fundamental mode does not depend on the specific shape of the bump.}
	\label{Fig: spiralAltBumps-diminish}
\end{figure}

In Fig.~\ref{Fig: spiral-1:x^2}, we compare the spiral of the fundamental mode due to a bump of a fixed size versus a bump whose height decreases proportionally to $1/L^2$, which is the same rate as the Regge-Wheeler potential drops to zero when the radial coordinate $r\rightarrow \infty$.  In both cases, the fundamental mode spirals away from the location of the unperturbed mode.  However, in the latter case the fundamental mode moves away at a much slower rate.  
In Fig.~\ref{Fig: spiralAltBumps-diminish}, we do the same procedure, but with the height decreasing proportionally to $1/L$, which corresponds to a bump that has a constant energy as $L$ increases.  In this figure, we have also included data for the fundamental mode of triangular and raised P\"oschl-Teller bumps to further illustrate that the specific shape of the bump does not change the behavior of the fundamental mode. 

\begin{figure}[th!]
	\begin{center}
		\includegraphics[height=8cm]{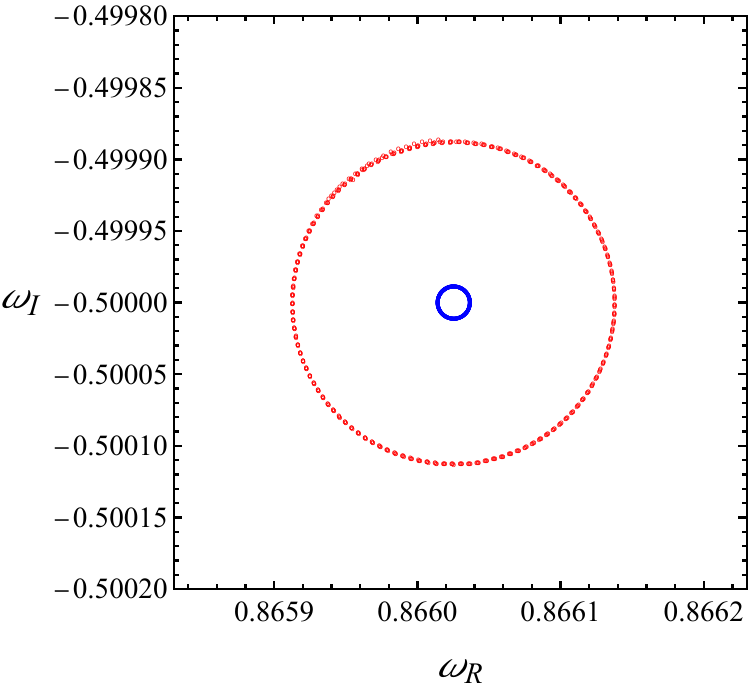}
        \includegraphics[height=8cm]{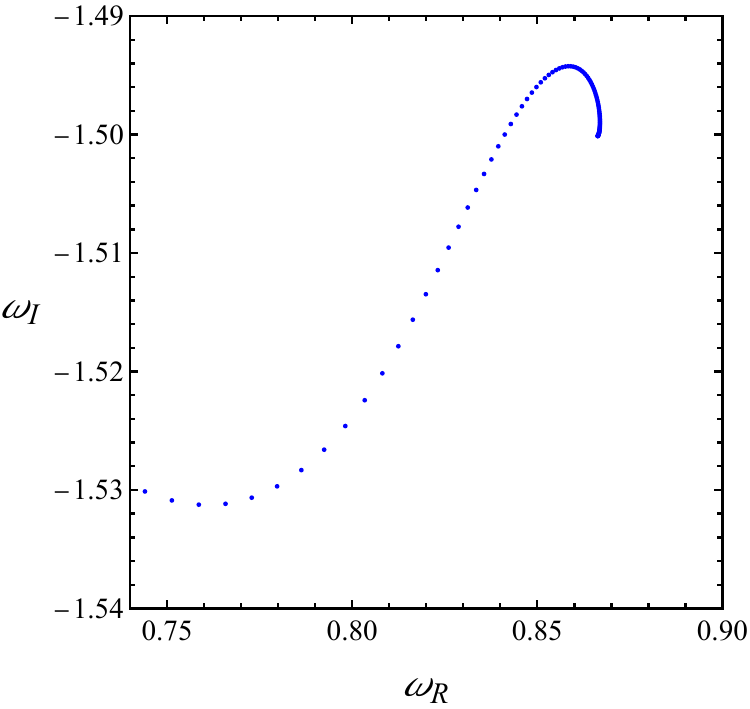}
	\end{center}
	\vspace{-0.7cm}
	\caption{\footnotesize The counter-clockwise motion of the fundamental QNM, left panel, and the outward spiral of the first overtone, right panel, are plotted in the frequency domain for the P\"oschl-Teller potential with a rectangular bump given in Eq.~\eqref{eq: VPTsquare}, where the height of the bump decreases exponentially. More specifically, we have  $\varepsilon=e^{-\kappa(L-L_0)}\varepsilon_0$, where $L_0=2\kappa^{-1}$ and we choose $\varepsilon_0=10^{-1} \kappa^2$ (blue dots in the left and right panels) and $\varepsilon_0=1 \kappa^2$ (red dots in the left panel).  For all cases, we have   $V=1 \kappa^2$,   $\sigma =10^{-4}\kappa^{-1}$.   For the unperturbed P\"oschl-Teller potential, the fundamental mode is located at $\omega_0\approx(\sqrt{3/4}-0.5 i) \kappa$ and  the first overtone is located at $\omega_1\approx(\sqrt{3/4}-1.5 i) \kappa$. In the right panel, the first overtone moves from right to left as we increase $L$.} 
	\label{Fig: spiral-ExpDecay}
\end{figure}

In Fig.~\ref{Fig: spiral-ExpDecay}, we show numerically the motion of the fundamental mode for the case where the rectangular bump diminishes exponentially in height proportional to $e^{-\kappa L}$, which is less than the rate $e^{-2\kappa L}$ at which the discontinuous P\"oschl-Teller potential \eqref{V_PT} decreases as $L \rightarrow \infty$.  Interestingly, for this case the fundamental mode is stable.  It rotates around the location of the fundamental mode of the unperturbed potential instead of spiraling outward or inward.  This is the critical rate of the decrease in the size of the perturbation in a P\"oschl-Teller potential, where a transition occurs from the outward spiral to the inward spiral.
%In Fig.~\ref{Fig: spiral-ExpDecay}, we show numerically the motion of the fundamental mode due to a barrier whose height decreases exponentially for two different initial barrier heights. It appears that the fundamental mode circles around the location of the fundamental mode of the original P\"oschl-Teller potential.  
This circular behavior can be explained analytically using the results we found in Eqs.~\eqref{Eq.:AnalyticSpiral}, \eqref{spiralEq}, and \eqref{Eq.:AnalyticSpiralRectangle}, where we have
\bqn
\delta \omega_n \propto  e^{2 i \omega_n L}.
\label{}
\eqn
To make the height of the bump diminish exponentially, we multiply $\epsilon$ in Eqs.~\eqref{Eq.:AnalyticSpiral} and \eqref{spiralEq} or $\varepsilon$ in Eq.~\eqref{Eq.:AnalyticSpiralRectangle} with $e^{-\kappa L}$ to get
\bqn
\delta \omega_n \propto
e^{-\kappa L}e^{2 i \omega_n L}.
\label{eq:exp-decrease-delta-w}
\eqn
Note that for a pure P\"oschl-Teller potential, the damping term of the QNM frequencies has the form
\bqn
\Im \omega_n = - \kappa \left(n+\frac{1}{2}\right).
\label{}
\eqn
Therefore, for for the fundamental mode with $n=0$, Eq.~\eqref{eq:exp-decrease-delta-w} reduces to $\delta \omega_0 \propto
e^{2 i \Re\omega_0 L}$.  In the complex plane, this represents a circular rotation, as shown in the left panel of Fig.~\ref{Fig: spiral-ExpDecay}, in the counterclockwise direction as $L$ increases.  For higher modes with $n\geq 1$,  Eq.~\eqref{eq:exp-decrease-delta-w} predicts an outward spiral.  Such an outward spiral is shown for the first overtone in the right panel of Fig.~\ref{Fig: spiral-ExpDecay}.

\section{Inward and outward spirals in a P\"oschl-Teller potential with a jump discontinuity}\label{sec6}

In this section, we explore another intriguing example by considering the following double-sided P\"oschl-Teller effective potential of the form
\begin{equation}
		\tilde{V}_\text{PT}(x) = \left\{ \begin{array}{lll}
			\frac{V_l}{\cosh ^2(\kappa_l x)}  &~~~~~& x \le L~\\   \\  
			\frac{V_r}{\cosh ^2(\kappa_r x)} &~~~~~& x > L~
		\end{array}
		\right. ,    
		\label{eq: VPTPT}
	\end{equation}
where the subscripts $l$ and $r$ refer to the left and right sides of the discontinuity at $L$, respectively.  
We have a jump discontinuity at $L$ when $V_l \neq V_r$.  
It is important to note that the effective potentials on both sides of the discontinuity can be realized by physically realistic systems, for instance, as effective potentials governed by specific compact objects, such as a central mass surrounded by an infinitely thin shell of matter. 
In this case, the magnitude of the perturbation is governed by the amount of mass in the shell.
 
%As the discontinuity moves from $-\infty$ to $+\infty$, the QNM spirals out of the location of the fundamental mode of the P\"oschl-Teller potential to the right of the discontinuity to the location of the fundamental mode of the potential to the left of the discontinuity
\begin{figure}[th!]
	\begin{center}
		\includegraphics[height=8cm]{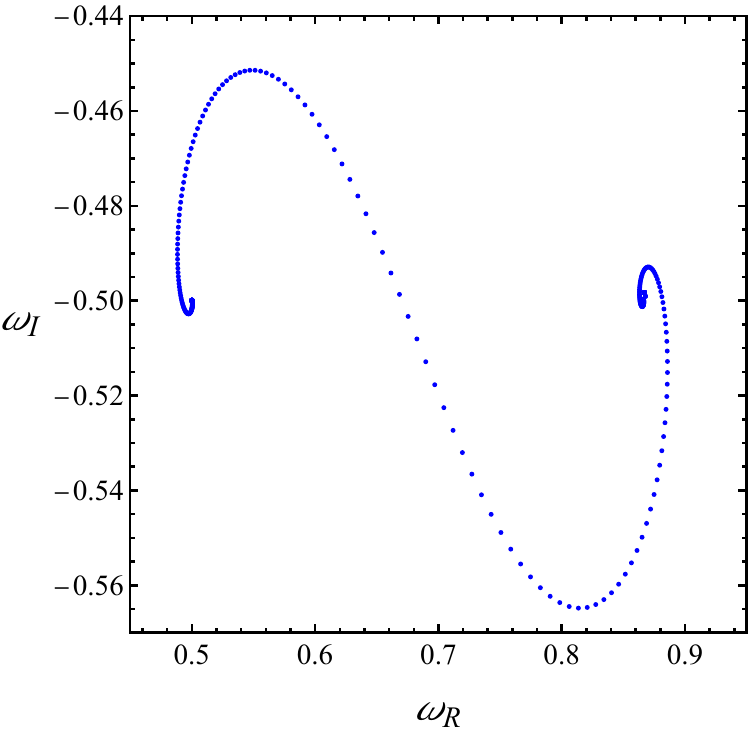}
	\end{center}
	\vspace{-0.7cm}
	\caption{\footnotesize The outspiral and inspiral of the fundamental QNM in the frequency domain is plotted for the effective potential in Eq. \eqref{eq: VPTPT}, where we choose $\kappa_l=\kappa_r=\kappa$, $V_l=1\kappa^{-1}$ and $V_r=0.5\kappa^{-1}$.  The data points move from left to right as $L$ increases from $-4\kappa^{-1}$ to $5\kappa^{-1}$.}
	\label{fig: PTPTSpiral}
\end{figure}
This case leads to an interesting scenario in which, by moving $L$ from $-\infty$ to $+\infty$, the fundamental mode initially spirals out starting from the location of the fundamental mode of the P\"oschl-Teller potential located in $x>L$ and then spirals towards the fundamental mode of the potential located in $x \le L$.  In contrast to all the previous examples provided above,  in this case the fundamental mode is stable as $L$ increases.  In Fig.~\ref{fig: PTPTSpiral}, we show the case where $\kappa_l=\kappa_r$ and $V_l > V_r$. 

The qualitative behavior of the spiral can be studied analytically for large values of $L$. The solutions to the wave Eq.~\eqref{pt_homo_eq} with the  effective potential
Eq.~\eqref{eq: VPTPT} are given in Eqs.~\eqref{eq: f-PT-delta} and \eqref{eq: g-PT-delta} for $x \le L$ and $x>L$ respectively.
After some simplification, the zeros of the Wronskian in Eq.~\eqref{eqWronskianCond} can be determined by solving
\bqn
&&
\beta _l \kappa _l \, _2F_1\left(1+\beta _l,\beta _l-\frac{i \omega }{\kappa _l};1-\frac{i \omega }{\kappa _l};-e^{2 L \kappa _l}\right) \, _2F_1\left(\beta _r,\beta _r-\frac{i \omega }{\kappa _r};1-\frac{i \omega }{\kappa _r};-e^{-2 L \kappa _r}\right)  \nonumber \\
&& +\beta _r \kappa _r \, _2F_1\left(\beta _l,\beta _l-\frac{i \omega }{\kappa _l};1-\frac{i \omega }{\kappa _l};-e^{2 L \kappa _l}\right) \, _2F_1\left(1+\beta _r,\beta _r-\frac{i \omega }{\kappa _r};1-\frac{i \omega }{\kappa _r};-e^{-2 L \kappa _r}\right)  \nonumber \\
&&+\frac{1}{2} \, _2F_1\left(\beta _l,\beta _l-\frac{i \omega }{\kappa _l};1-\frac{i \omega }{\kappa _l};-e^{2 L \kappa _l}\right) \, _2F_1\left(\beta _r,\beta _r-\frac{i \omega }{\kappa _r};1-\frac{i \omega }{\kappa _r};-e^{-2 L \kappa _r}\right) \nonumber \\
&& ~~~~ \times \left\{ \beta _l \kappa _l \left[\tanh \left(L \kappa _l\right]-1\right)-\beta _r \kappa _r \left[\tanh \left(L \kappa _r\right)+1\right]-2 i \omega \right\} =0.
\label{eq: PT-PT-spiral}
\eqn
Considering the case where $\kappa_l=\kappa_r=\kappa$, for large values of $z=e^{2 \kappa  L}$, we can use the series expansion of the above equation using Eqs.~\eqref{eq: Series1} and \eqref{eq: Series2} to obtain
\begin{equation}
-i \frac{n!}{\kappa} \delta \omega_n \approx \frac{1}{\Gamma \left(\beta_l -\frac{i \omega}{\kappa }\right)}=
\frac{i \sin \left(\pi  \beta _l\right) \left(\beta _l-\beta _r\right) \left(\beta _l+\beta _r-1\right) \text{csch}\left(\frac{\pi  \omega_n }{\kappa}\right) \Gamma \left(1-\beta _l-\frac{i \omega_n }{\kappa}\right) }{\Gamma \left(1-\frac{i \omega_n }{\kappa }\right) \Gamma \left(2-\frac{i \omega_n }{\kappa}\right)}
e^{-2 \kappa L}e^{2 i  \omega_n L }.
\label{eq: In-Out-spiral1}
\end{equation}
This is a limiting case of the discontinuous P\"ochl-Teller potential with a delta-function, discussed in Sec.~\ref{sec31}, as the strength of the delta-function goes to zero. Note that Eq.~\eqref{Eq.:AnalyticSpiral-withDisontinuity}  reduces to Eq.~\eqref{eq: In-Out-spiral1} when $\epsilon=0$.

The discontinuity can also be moved in the negative direction, where $L\rightarrow -\infty$.  In this case, we can Taylor expand Eq.~\eqref{eq: PT-PT-spiral} for large values of $z=e^{-2\kappa L}$.  This gives us  
\bqn
-i \frac{n!}{\kappa} \delta \omega_n \approx \frac{1}{\Gamma \left(\beta_r -\frac{i \omega}{\kappa }\right)}=-
\frac{i \sin \left(\pi  \beta _r\right)  \left(\beta _l-\beta _r\right) \left(\beta _l+\beta _r-1\right)   \text{csch}\left(\frac{\pi  \omega_n }{\kappa }\right)\Gamma \left(1-\beta _r-\frac{i \omega_n }{\kappa }\right)}{  \Gamma \left(1-\frac{i \omega_n }{\kappa }\right) \Gamma \left(2-\frac{i \omega_n }{\kappa }\right)}
e^{2 \kappa L}e^{-2 i  \omega_n L }.
\label{eq: In-Out-spiral2}
\eqn
In Fig.~\ref{fig: PTPTSpiral}, we show the outspiral and inspiral of the fundamental QNM as $L$ moves from large negative to large positive values.  The outward spiral followed by an inward spiral shown in  Fig.~\ref{fig: PTPTSpiral} agrees well with the analytic results \eqref{eq: In-Out-spiral1} and \eqref{eq: In-Out-spiral2}.

Note that Eq.~\eqref{eq: In-Out-spiral1} contains the factor $e^{-2\kappa L}$, which does not appear in Eqs.~\eqref{Eq.:AnalyticSpiral}, \eqref{spiralEq}, and \eqref{Eq.:AnalyticSpiralRectangle}.  This factor is responsible for the stability of the fundamental mode $\omega_0$.  This is because, according to Eq.~\eqref{V_PT-QNM}, $e^{-2\kappa L} e^{2i\omega_0 L}=e^{-\kappa L}e^{2 i L \Re \omega_0}\rightarrow 0$ as $L$ increases.  This clearly represents an inward spiral.  

The factor $e^{-2\kappa L}$ appears because the size of the discontinuity at large $L$ is 
\bqn
\frac{V_l}{\cosh ^2(\kappa L)}-\frac{V_r}{\cosh ^2(\kappa L)} \sim e^{-2\kappa L} (V_l-V_r),
\eqn
which diminishes proportionally to $e^{-2\kappa L}$ with increasing $L$.    From the analytic results of Eqs.~\eqref{Eq.:AnalyticSpiral} and \eqref{spiralEq}, it is easy to see that one can achieve an inward spiral by decreasing the strength of the delta-function, $\epsilon$, proportionally to $e^{-2\kappa L}$.  In the case of the rectangular perturbation in Eq.~\eqref{Eq.:AnalyticSpiralRectangle}, the inward spiral can be achieved by decreasing the height ($\varepsilon$), the width ($\sigma$), or the area of the rectangle proportionally to $e^{-2\kappa L}$.

\section{The motion of the fundamental mode in a perturbed Regge-Wheeler potential}\label{sec7}

Now, we turn to more realistic scenarios by studying the motion of the fundamental mode of a perturbed Regge-Wheeler effective potential as the perturbation moves away from the central black hole.  
Specifically, we explore two cases.  
In the first case, we truncate the Regge-Wheeler potential in Eq.~\eqref{V_RW}, where we set the potential equal to zero for $r> L$.  In the second case, which we call double-sided, we replace the black hole mass $M$ in Eq.~\eqref{V_RW} with $M+\delta M$ for $r> L$.  This models a black hole surrounded by a thin spherical matter shell located at $r=L$.\footnote{To be more accurate, a redshift factor should be included, as done in \cite{Laeuger:2025zgb}. For the $\delta M$ used in this paper, the redshift contribution is too small (approximately a $1\%$ correction) to make a qualitative difference, so we omit it for simplicity.}  

To calculate and track the fundamental mode, we use Leaver's continued fraction method~\cite{agr-qnm-continued-fraction-01}.  In particular, since our effective potential exhibits a discontinuity, to evaluate the fundamental mode, we follow the technique laid out in \cite{Li:2026xnr}. 
This extension of Leaver's method involves expanding the wavefunction about the location of the discontinuity, where the solutions on either side of the discontinuity are matched via a junction condition~\eqref{pt_Wronskian}. %To maintain accuracy, we found it was necessary to increase the depth of the continued fractions as the perturbation approached the horizon.

\begin{figure}[th!]
	\begin{center}
		\includegraphics[height=8cm]{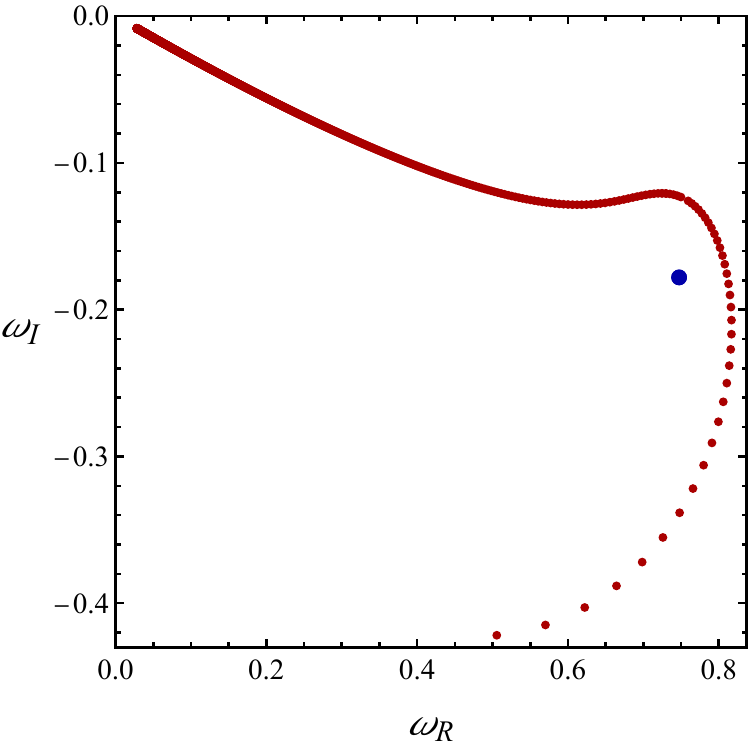}
	\end{center}
	\vspace{-0.7cm}
	\caption{\footnotesize The motion of the fundamental QNM in the frequency domain (red dots moving upward) is plotted for the truncated Regge-Wheeler potential, for gravitational perturbations with multipole number $\ell=2$, as the location of the truncation moves away from the central black hole.  The blue dot represents the location of the fundamental mode of the unperturbed Regge-Wheeler potential.  }
	\label{Fig: RW}
\end{figure}

In Fig.~\ref{Fig: RW}, we show the motion of the fundamental mode in the truncated Regge-Wheeler potential as $L$ increases.  For small values of $L$, the effective potential is significantly deformed and the fundamental mode is initially far from the unperturbed value, shown in blue.  As $L$ increases, the mode approaches the unperturbed value. As $L$ continues to increase, it moves away. 

\begin{figure}[th!]
	\begin{center}
		\includegraphics[height=8cm]{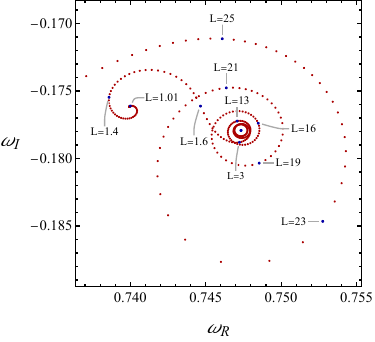}
        \includegraphics[height=8.cm]{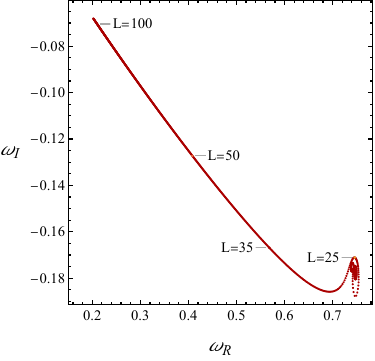}
	\end{center}
	\vspace{-0.7cm}
	\caption{\footnotesize The motion of the fundamental QNM in the frequency domain (red dots spiraling counter-clockwise in the left panel, before moving to the left in the right panel) is plotted for the double-sided Regge-Wheeler potential, for gravitational perturbations with multipole number $\ell=2$, as the location of the discontinuity, $L$ in units of $r_h$, moves away from the central black hole.  The blue bold dot in the center of the left panel represents the location of the fundamental mode of the unperturbed Regge-Wheeler potential with mass $M$.  The blue bold dot at $L=1.01 r_h$ of the left panel represents the location of the fundamental mode of the unperturbed Regge-Wheeler potential with mass $M+\delta M$. The discontinuity is caused by an increase in the mass by $\delta M=0.005 M$.}
	\label{Fig: RW-ds}
\end{figure}

In Fig.~\ref{Fig: RW-ds}, we show the motion of the fundamental mode in the double-sided Regge-Wheeler potential as $L$ changes.  
For large $L$, the fundamental mode is far from the unperturbed value. We conjecture that this is due to the appearance of echo modes when $L$ is sufficiently far from the black hole, as explained in \cite{agr-qnm-instability-65}. As $L$ decreases, the fundamental mode moves closer to the fundamental mode of the unperturbed black hole with mass $M$.  As $L$ continues to decrease, it spirals around this mode until it begins to spiral out (around $L=3 r_h$ in the diagram) and moves toward the fundamental mode for the black hole with mass $M + \delta M$.
At  $L = 2(M + \delta M) =1.01 r_h$ the event horizon discontinuously jumps from $2M$ to $2(M + \delta M)$.  To avoid this nonphysical behavior, starting at $L = 1.5 r_h$, we slowly increase the mass of the black hole from $M$ to $M + \delta M$ as $L$ decreases to $1.01 r_h$.  As a result, the fundamental mode approaches that of an unperturbed black hole with mass $M + \delta M$.  Notice, this behavior for small $L$, in the radial coordinate, resembles the behavior for the double sided P\"oschl-Teller example as $L\rightarrow -\infty$ in the tortoise coordinate (Fig.~\ref{fig: PTPTSpiral}).  This is due to the fact that, in the tortoise coordinate, both the P\"oschl-Teller and the Regge-Wheeler potential decrease exponentially as we approach the event horizon.  If we were to use the tortoise coordinate in the calculation for Fig.~\ref{Fig: RW-ds}, we could avoid changing the mass, since the two horizons (corresponding to $M$ and $M+\delta M$) would both naturally approach each other as $L$ approaches the event horizon at $-\infty$.


\section{Concluding remarks}\label{secConclude}

In this work, we investigated the sensitivity of the fundamental QNM to ultraviolet perturbations of the effective potential by analyzing a variety of analytically tractable and numerically accessible scenarios.
In most cases, the motion of the mode in the complex-frequency plane exhibits a spiral structure as the location of the perturbation moves away from the central object.
The analytic results derived in the present study agree well with the numerical calculations and provide a qualitative understanding of the underlying mechanisms responsible for the spectral instability of the fundamental mode.

The instability of the fundamental mode is governed primarily by the effective size of the perturbation rather than by its detailed shape. 
In particular, the instability may persist even in the limit where the width of the perturbative bump approaches zero, as a delta-function perturbation still induces an outward spiral provided that the integrated strength of the perturbation is not zero. 
This result indicates that the integrated strength (or effective size) of the perturbation provides a more physically meaningful characterization than its peak amplitude, or its width, alone. 
Accordingly, while the quantitative features of the spiral trajectories depend on the adopted normalization of the perturbation strength as the perturbation is displaced, the underlying instability mechanism remains qualitatively unchanged. 
In a discontinuous P\"ochl-Teller potential, the fundamental mode is stable in the sense that an inward spiral is observed.
While a discontinuous P\"ochl-Teller potential alone induces an inward spiral, the combination of a discontinuity and a localized perturbation can give rise to an interplay between inward and outward spiral motions.
We show that the stability of the fundamental mode can be affected if the magnitude of the perturbation decreases as it moves away from the central object. 
A slowly decreasing perturbation continues to induce an outward spiral, while a faster decrease leads to an inward spiral. 
In fact, the rate of decrease can be tuned to induce a circular motion. 
It is also shown that in a double-sided P\"oschl-Teller potential, the fundamental mode follows an outspiral-inspiral structure, where the fundamental mode interpolates between the spectra associated with two distinct asymptotic effective potentials. 
The inspiral occurs because the size of the discontinuity in this case naturally decreases at a fast enough rate, matching the rate of decrease in the unperturbed P\"oschl-Teller potential.

We also explored the motion of the fundamental mode in a perturbed Regge-Wheeler potential containing a jump discontinuity associated with a thin matter shell surrounding the black hole.
As the radius of the shell increases, the mode initially spirals outward from the location associated with a black hole of mass $M+\delta M$, followed by an inward spiral toward the mode of the original black hole of mass $M$, before spiraling outward again.
This behavior qualitatively resembles the spiral structure observed in the double-sided P\"oschl-Teller potential and suggests that the mechanisms identified in the analytically tractable models persist in more realistic black hole effective potentials.

The present analysis has been restricted to localized perturbations that preserve the spherical symmetry of the background spacetime, allowing the perturbation equations to be reduced to a one-dimensional radial master equation. 
Although this framework captures the response of a specific angular sector and provides a systematic understanding of the sensitivity of QNMs to localized ultraviolet perturbations, realistic astrophysical environments generally involve non-spherical matter distributions and dynamical backgrounds.

Overall, the present results indicate that the spectral instability of low-lying black hole modes is considerably more intricate than previously anticipated. 
Although the fundamental mode is substantially more robust than higher overtones, its behavior remains sensitive to the localization, integrated strength, and asymptotic scaling of localized ultraviolet perturbations. 
While the present analysis is restricted to perturbations that preserve the spherical symmetry of the background spacetime and are introduced phenomenologically at the level of the effective potential, it identifies robust mechanisms governing how localized perturbations influence the quasinormal spectrum. 
Extending the present framework to more realistic astrophysical scenarios involving non-spherical matter distributions or dynamical backgrounds constitutes an important direction for future work. 
Such studies will be essential for assessing the implications of the present findings for black hole spectroscopy and the interpretation of future ringdown observations.

\section{Acknowledgments}
We acknowledge the financial support from Brazilian agencies 
Funda\c{c}\~ao de Amparo \`a Pesquisa do Estado de S\~ao Paulo (FAPESP), 
Funda\c{c}\~ao de Amparo \`a Pesquisa do Estado do Rio de Janeiro (FAPERJ), 
Conselho Nacional de Desenvolvimento Cient\'{\i}fico e Tecnol\'ogico (CNPq), 
and Coordena\c{c}\~ao de Aperfei\c{c}oamento de Pessoal de N\'ivel Superior (CAPES).
This work is also supported by the National Natural Science Foundation of China (NSFC).
A part of this work was developed under the project Institutos Nacionais de Ci\^{e}ncias e Tecnologia - F\'isica Nuclear e Aplica\c{c}\~{o}es (INCT/FNA) Proc. No. 408419/2024-5.
This research is also supported by the Center for Scientific Computing (NCC/GridUNESP) of S\~ao Paulo State University (UNESP).

\bibliographystyle{h-physrev}
\bibliography{references_qian, references_daghigh}

\end{document}